\documentclass[useAMS,usenatbib]{mn2e}

\usepackage{aas_macros}

\usepackage{graphicx}
\mathchardef\mhyphen="2D

\title[Anti-truncated light profiles in STAGES spirals]{Anti-truncated stellar light profiles in the outer
regions of STAGES spiral galaxies: bulge or disc related?}

\author[D.~T.~Maltby et al.]
{David~T.~Maltby$^{1}$\thanks{E-mail: ppxdtm@nottingham.ac.uk},
 Carlos~Hoyos$^{1}$,
 Meghan~E.~Gray$^{1}$,
 Alfonso~Arag{\'o}n-Salamanca$^{1}$,
 \newauthor Christian~Wolf$^{2}$, \\
$^{1}$School of Physics and Astronomy, The University of Nottingham, University Park, Nottingham, NG7 2RD, UK. \\
$^{2}$Department of Physics, Denys Wilkinson Building, University of Oxford, Keble Road, Oxford, OX1 3RH, UK.}

\begin{document}

\date{Accepted 2011 November 16. Received 2011 November 15; in original form 2011 August 25}

\pagerange{\pageref{firstpage}--\pageref{lastpage}} \pubyear{0000}

\maketitle

\label{firstpage}


\begin{abstract}

We present a comparison of azimuthally averaged radial surface brightness $\mu(r)$ profiles and analytical
bulge--disc decompositions (de Vaucouleurs, $r^{1/4}$ bulge plus exponential disc) for spiral galaxies
using {\em Hubble Space Telescope}/Advanced Camera for Surveys $V$-band imaging from the Space Telescope
A901/2 Galaxy Evolution Survey (STAGES). In the established classification scheme, antitruncated $\mu(r)$
profiles (Type~III) have a broken exponential disc with a shallower region beyond the break radius
$r_{\rm brk}$. The excess light at large radii ($r>r_{\rm brk}$) can either be caused by an outer
exponential disc (Type~III-d) or an extended spheroidal component (Type~III-s). Using our comparisons, we
determine the contribution of bulge light at $r>r_{\rm brk}$ for a large sample of $78$ (barred/unbarred,
Sa-Sd) spiral galaxies with outer disc antitruncations ($\mu_{\rm brk} > 24\,\rm mag\,arcsec^{-2}$). In the
majority of cases ($\sim85$ per cent), evidence indicates that excess light at $r>r_{\rm brk}$ is related to
an outer shallow disc (Type~III-d). Here, the contribution of bulge light at $r>r_{\rm brk}$ is either
negligible ($\sim70$ per cent) or too little to explain the antitruncation ($\sim15$ per cent). However in
the latter cases, bulge light can affect the measured disc properties (e.g. $\mu_{\rm brk}$, outer
scalelength). In the remaining cases ($\sim15$ per cent), light at $r>r_{\rm brk}$ is dominated by the bulge
(Type~III-s). Here, for most cases the bulge profile dominates at all radii and only occasionally ($3$
galaxies, $\sim5$ per cent) extends beyond that of a dominant disc and explains the excess light at
$r>r_{\rm brk}$. We thus conclude that in the vast majority of cases antitruncated outer discs cannot be
explained by bulge light and thus remain a pure disc phenomenon.

\end{abstract}


\section[]{Introduction}

\label{Introduction}

The light profiles of spiral galaxies consist of two principal components: an inner, bulge-dominated
component; and an outer exponentially declining disc with some minor deviations related to spiral arms
\citep{deVaucouleurs:1959, Freeman:1970}. However, since \cite{vanderKruit:1979} we have known that this
`classical' picture fails for most spiral galaxies, particularly at the faint surface brightness $\mu$ of
the outer stellar disc. We now know that most disc profiles are best described by a two slope model (broken
exponential), characterised by an inner and outer exponential scalelength separated by a relatively well
defined break radius $r_{\rm brk}$ \citep{Pohlen_etal:2002}. Many studies have now reported (mainly using
surface photometry) the existence of broken exponential discs, or {\em truncations}, in spiral galaxies in
both the local \citep{Pohlen_etal:2002, Pohlen_etal:2007, Pohlen_Trujillo:2006, Bakos_etal:2008,
Erwin_etal:2008, Gutierrez_etal:2011, Maltby_etal:2011} and distant $z<1$ Universe \citep{Perez:2004,
Trujillo_Pohlen:2005, Azzollini_etal:2008}. Broken exponential discs have also been reported using resolved
star counts on some nearby galaxies \citep{Ibata_etal:2005,Ferguson_etal:2007}.

These studies have resulted in a comprehensive classification scheme for disc galaxies based on break
features in the outer disc component of their radial $\mu$ profiles \citep[see e.g.][]{Erwin_etal:2005,
Erwin_etal:2008, Pohlen_Trujillo:2006}. This scheme consists of three broad profile types (Type I, II and
III): Type I (no break) -- a single exponential disc extending out to several scalelengths
\citep[e.g.][]{BlandHawthorn_etal:2005}; Type II (down-bending break, {\em truncation}) -- a broken
exponential disc with a shallow inner and steeper outer region \citep{vanderKruit:1979,Pohlen_etal:2002};
Type III (up-bending break, {\em antitruncation}) -- a broken exponential disc with a shallower region beyond
the break radius $r_{\rm brk}$ \citep{Erwin_etal:2005}.


In the classical picture (simple bulge and disc), the de Vaucouleurs, $r^{1/4}$ bulge profile dominates in
the centre while the exponential disc dominates at larger radii. However, theoretically the $r^{1/4}$
profile would dominate again at some low surface brightness. \cite{Erwin_etal:2005} suggest that in some
Type~III profiles (up-bending breaks), the excess light beyond the break radius $r_{\rm brk}$ could
be attributed to light from the spheroidal bulge or halo extending beyond the end of the disc. Consequently,
Type~III profiles can be separated into two distinct sub-classes depending on whether the outer profile
$r>r_{\rm brk}$ is dominated by a disc (Type~III-d) or spheroidal component (Type~III-s).
\cite{Erwin_etal:2005} also propose that antitruncations with a smooth gradual transition and outer isophotes
that are progressively rounder than that of the main disc, suggest an inclined disc embedded in a more
spheroidal outer region such as an extended bulge or halo (i.e. Type~III-s). Using this `ellipse' method,
previous works \citep{Erwin_etal:2005, Erwin_etal:2008, Gutierrez_etal:2011} have found that $\sim40$ per
cent of their Type~III profiles are Type~III-s. 


However, the ellipse method is limited for face-on discs and cases where the outer/inner disc may have
different orientations and axis ratios. In these instances, bulge--disc \mbox{(B-D)} decomposition
\citep[e.g.][]{Allen_etal:2006} provides a useful tool to determine the contribution of the two major
structural components (bulge and disc) to the galaxy's light distribution and should provide more conclusive
evidence. The aim of this work is to use B-D decomposition on a large sample of $78$ outer disc
antitruncations from the Space Telescope A901/2 Galaxy Evolution Survey \citep[STAGES;][]{Gray_etal:2009} and
assess the fraction of Type~III profiles that show evidence for the excess light at large radii being caused
or affected by the spheroidal component. This work builds on previous studies by using an improved method for
the classification of Type III-s/III-d profiles (especially for face-on discs) and by using a larger more
representative sample spanning the range of spiral morphologies.

Throughout this paper, we adopt a cosmology of $H_0=70\,{\rm kms^{-1}Mpc^{-1}}$, $\Omega_\Lambda=0.7$, and
$\Omega_m=0.3$, and use AB magnitudes unless stated otherwise.

\section[]{Data and Sample Selection}

\label{Data and Sample Selection}

This work is entirely based on the STAGES data published by \cite{Gray_etal:2009}. STAGES is an extensive
multiwavelength survey targeting the Abell 901/2 multicluster system ($z\sim0.167$) and covering a wide
range of galaxy environments. {\em Hubble Space Telescope} ({\em HST})/Advanced Camera for Surveys (ACS)
$V$-band (F606W) imaging covering the full $0.5^{\circ}\times0.5^{\circ}$ ($\sim5\times5\,{\rm Mpc^2}$) span
of the multicluster system is complemented by extensive observations including photometric redshifts from
\cite{Wolf_etal:2003}. All imaging and data are publicly
available\footnote{http://www.nottingham.ac.uk/astronomy/stages}. In addition to this, all galaxies with
apparent $R_{\rm vega} < 23.5\,{\rm mag}$ and $z_{\rm phot} < 0.4$ (5090 galaxies) were visually
classified by seven members of the STAGES team into the Hubble types (E, S0, Sa, Sb, Sc, Sd, Irr)
and their intermediate classes (Gray et al. in prep.).


Our galaxy sample is drawn from \cite{Maltby_etal:2011}. This consists of a large, mass-limited
($M_* > 10^9\,\rm M_\odot$), visually classified (Sa-Sdm, from Gray et al. in prep) sample of $327$
face-on to intermediate inclined ($i< 60\,\rm degrees$) spiral galaxies from both the field and cluster
environments. However, we remove two galaxies for which B-D decomposition fails ($N_{\rm tot} = 325$). The
$182$ cluster spirals are at a redshift of $z_{\rm cl} = 0.167$ and the $143$ field spirals span a redshift
range of $0.05 < z_{\rm phot} < 0.30$. \cite{Maltby_etal:2011} analysed the $\mu(r)$ profiles for these
galaxies in order to identify broken exponentials in the outer stellar disc $\mu > 24\,\rm mag\,arcsec^{-2}$
(their criteria for selecting intrinsically similar outer breaks). Three independent assessors agreed on a
sub-sample of $78$ antitruncated (Type~III) outer $\mu$ profiles ($\mu_{\rm brk} > 24\,\rm mag\,arcsec^{-2}$).
We shall use both this Type~III sub-sample and the total sample in this work.

\section{Methodology}

\label{Methodology}

For each galaxy in our sample, we perform a two-dimensional B-D decomposition based on a two component
galaxy model comprising of a de Vaucouleurs ($r^{1/4}$) bulge and a single exponential disc. Decompositions
were carried out on the STAGES $V$-band imaging using the {\sc galfit} code \citep{Peng_etal:2002} and the
method of \cite{Hoyos_etal:2011} adapted to perform two component fits. Several measurable properties are
produced for each galaxy including: position [$x$,$y$], effective radii, total magnitudes, axis ratios,
position angles for the bulge and disc components, and a sky-level estimation. 

B-D decomposition can be sensitive to the initial conditions used to search the B-D parameter space (e.g.
initial guess for bulge-to-disc ratio $B/D$). Therefore, we perform two runs of the B-D decomposition with
different initial conditions from the two extremes. One run starting from a bulge-dominated system
($B/D = 9$), and the other run starting from a disc-dominated system ($B/D = 1/9$). Comparison of these runs
(hereafter, Run~$1$ and Run~$2$) allows for an assessment of the uniqueness/stability of B-D decomposition
on a galaxy-galaxy basis. In the vast majority of cases ($\sim85$ per cent) the results were effectively the
same, $\sim70$ per cent being exactly the same and $\sim15$ per cent having only minor differences that do
not affect our analysis ($B/D$ the same within $\sim10$ per cent). In a few cases ($\sim10$ per cent) the
decomposition was catastrophically unstable with Run 1/2 yielding both bulge- and disc-dominated systems. The
remaining cases ($\sim5$ per cent) showed moderate instabilities great enough to affect the assessment of
bulge light in the outer regions of the galaxy. These stability fractions are the same in our
Type~III sub-sample. The unstable solutions are mainly driven by differences in the sky level determined
during the decomposition. However, the overall conclusions of this work are not affected by these unstable
solutions. 


For each galaxy, we also use the {\sc iraf} task {\em ellipse} (STSDAS package - version 2.12.2) in order to
obtain azimuthally averaged radial $\mu(r)$ profiles from the STAGES $V$-band imaging, see
\cite{Maltby_etal:2011} for full details of the {\em ellipse} fitting method used. For all our {\em ellipse}
fits the galaxy centre is fixed (all isophotes have a common centre) using the galaxy centre determined
during B-D decomposition. The bad pixel masks of \cite{Gray_etal:2009} are also used to remove everything not
associated with the galaxy itself from the isophotal fit (e.g. light from companion galaxies).


Using a similar procedure to previous works \citep{Pohlen_Trujillo:2006,Erwin_etal:2008}, we fit two
different sets of ellipses to each galaxy image. The first is a free parameter fit (fixed centre, free
ellipticity $e$ and position angle ${\it PA}$) and tends to follow morphological features such as bars and
spiral arms. Consequently, these free-fits are not suitable for the characterisation of the underlying outer
disc, however they may be used to determine the $e$ and ${\it PA}$ of the outer disc component, see
\cite{Maltby_etal:2011} for further details. A fixed-parameter fit (fixed centre, $e$ and $\it{PA}$ using the
$e$ and $\it PA$ determined for the outer disc) is then used to produce our final measured $\mu(r)$
profiles. The necessary sky subtraction is then performed using the sky-level estimates generated during B-D
decomposition. Please note that these sky-values sometimes differ slightly from those of
\cite{Maltby_etal:2011}.


Analogous fixed-parameter fits (using the $e$ and $\it PA$ determined for the outer disc) are also carried
out on the disc-residual images (ACS image - bulge-only model) resulting in a measured $\mu$ profile for the
disc component $\mu_{\rm Disc}(r)$. We also obtain azimuthally averaged radial $\mu$ profiles for the
decomposed B-D model using the same fixed-parameter ellipses (isophotes) as in the other profiles. These
yield separate radial $\mu$ profiles for both the disc and bulge model along the semi-major axis of the outer
disc.


All resultant $\mu$ profiles are corrected for Galactic foreground extinction, individual galaxy inclination
$i$, and surface brightness dimming (the $\mu$ profiles of field galaxies, $0.05 < z_{\rm phot} < 0.30$, are
corrected to the redshift of the cluster $z = 0.167$). Full details of the fitting procedure, subsequent
photometric calibration, and an estimation of the error in the sky subtraction ($\pm0.18$ counts) can be
found in \cite{Maltby_etal:2011}.

\section[]{Results}

\label{Results}

\begin{figure}
\centering
\includegraphics[width=0.2\textwidth]{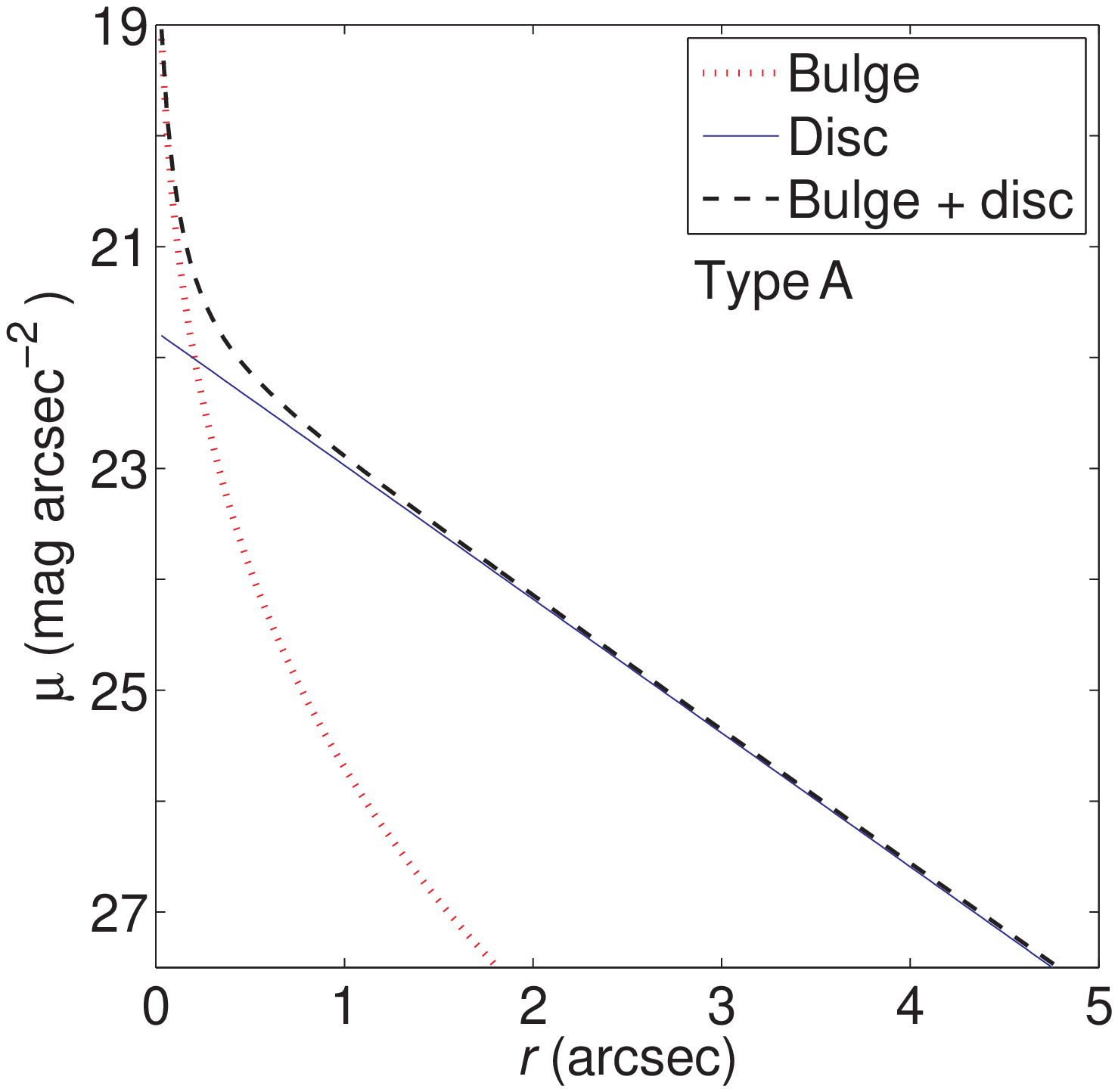}
\includegraphics[width=0.2\textwidth]{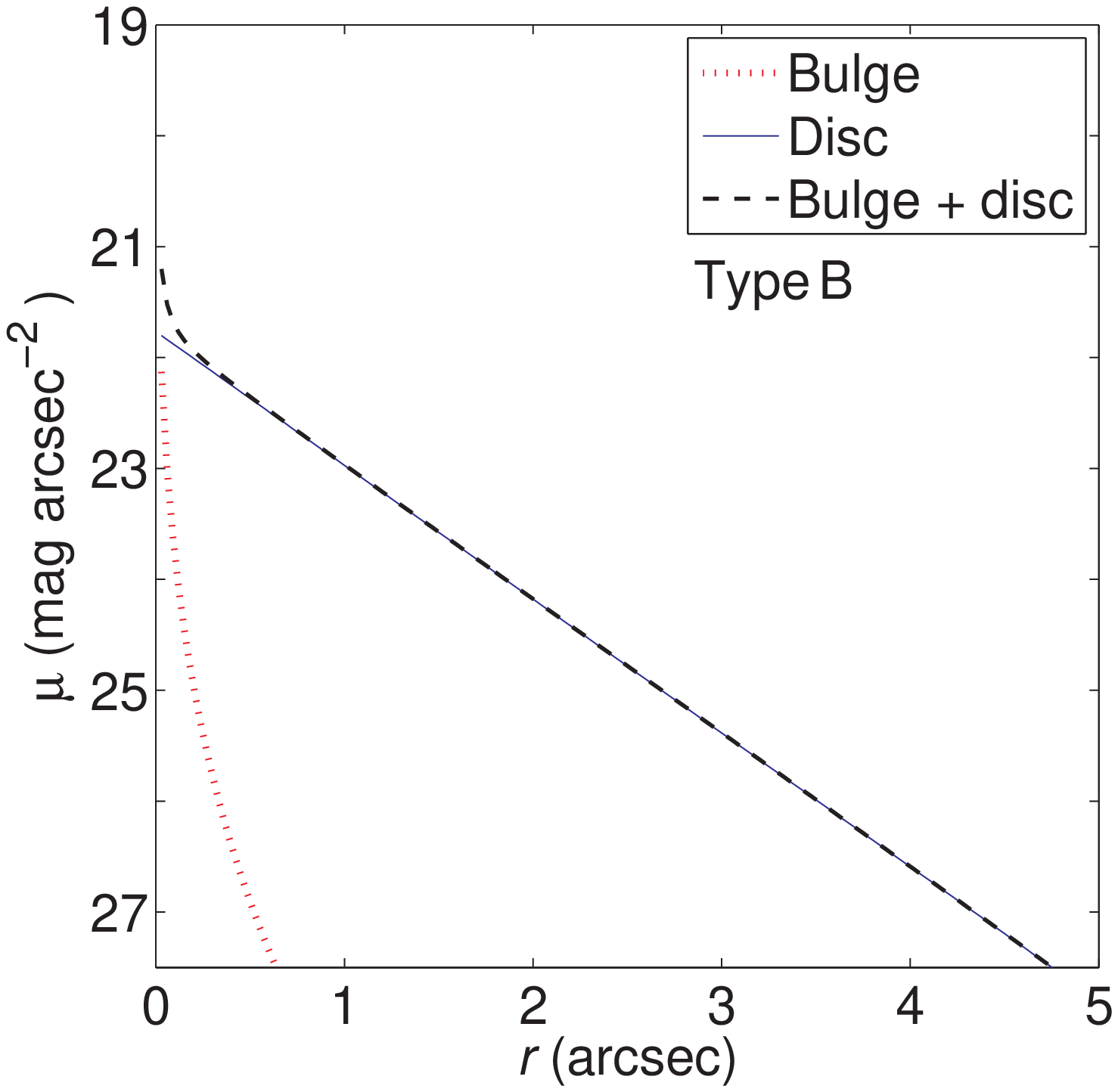}\\
\includegraphics[width=0.2\textwidth]{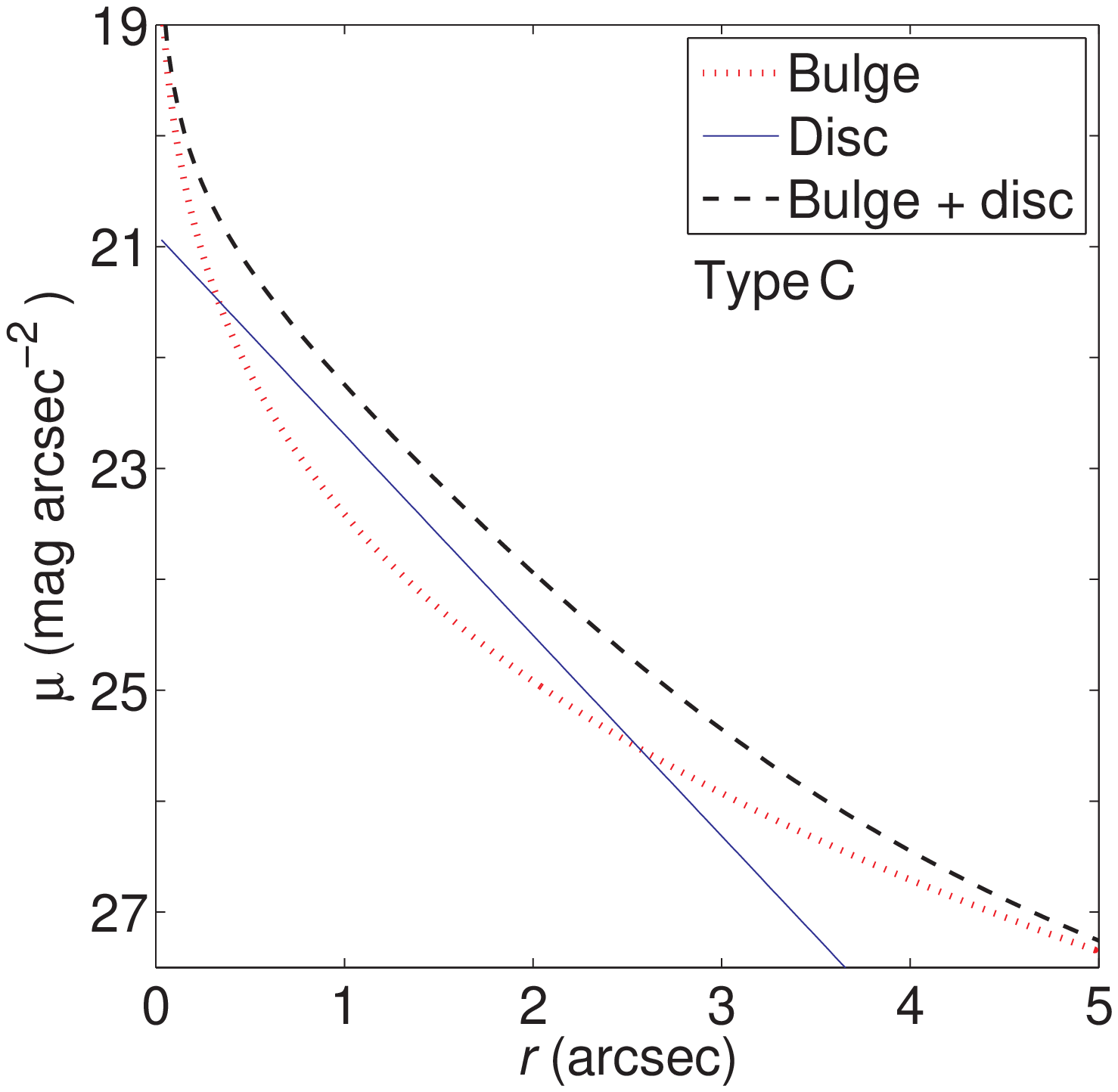}
\includegraphics[width=0.2\textwidth]{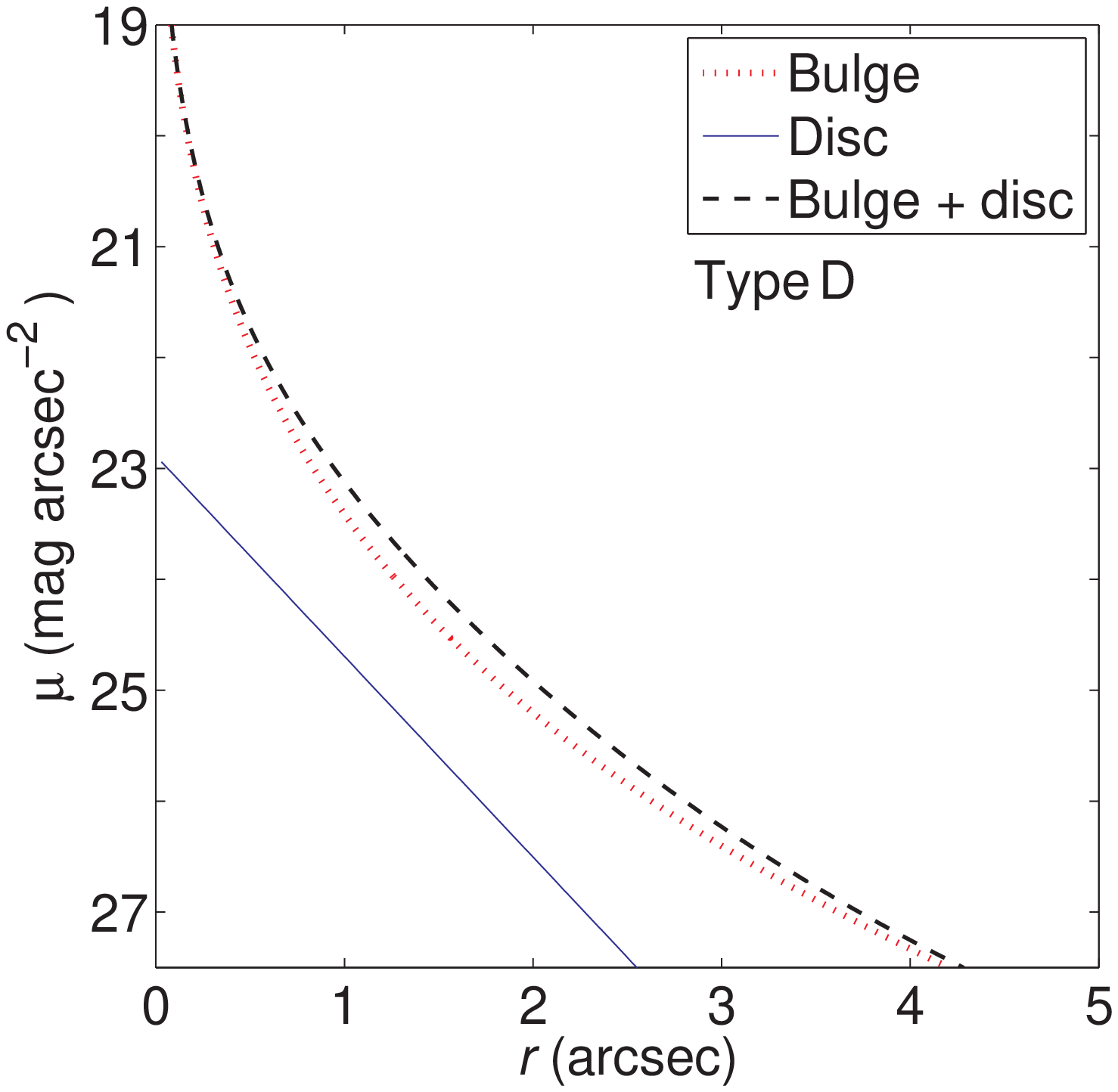}\\
\includegraphics[width=0.2\textwidth]{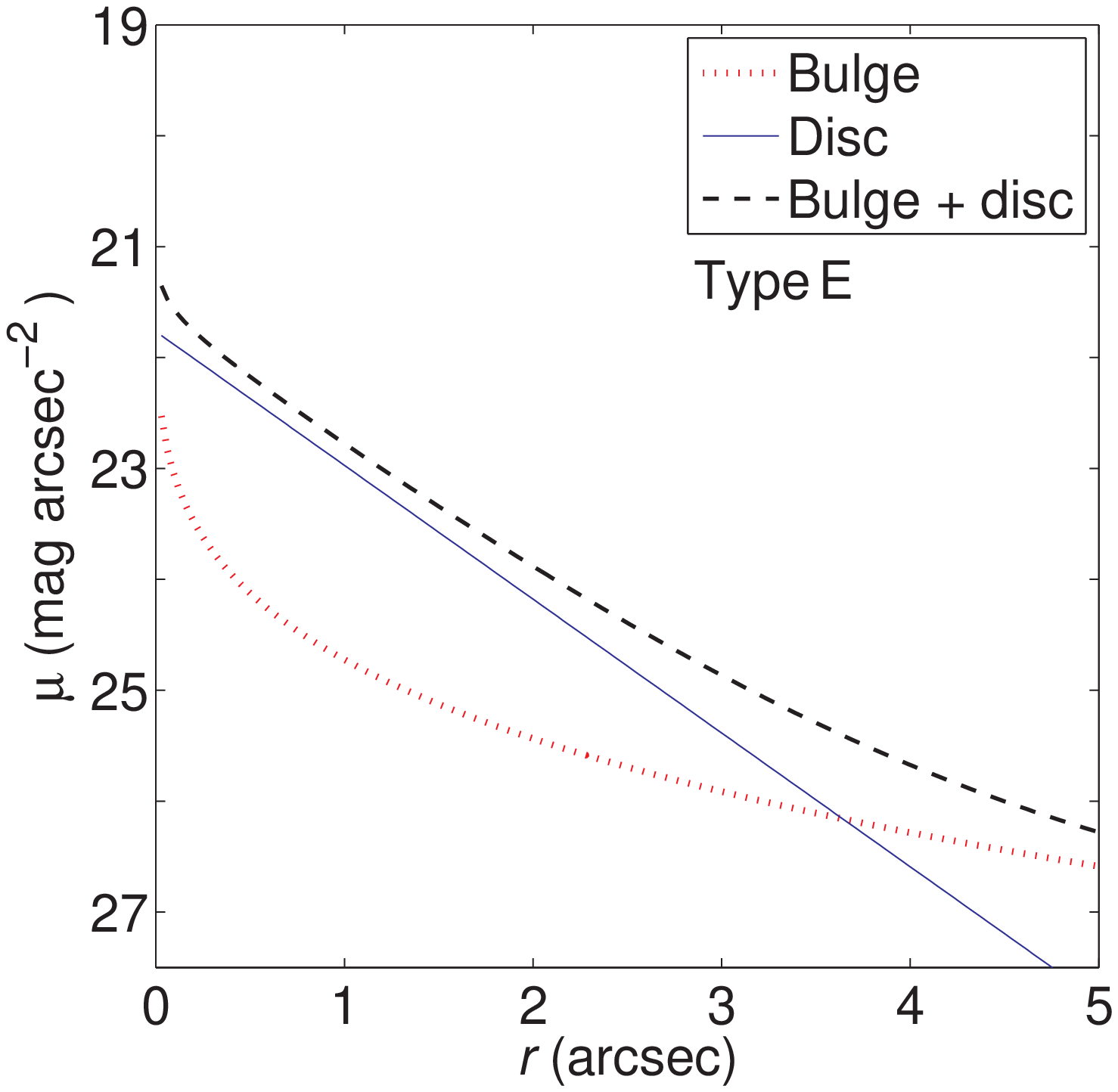}
\includegraphics[width=0.2\textwidth]{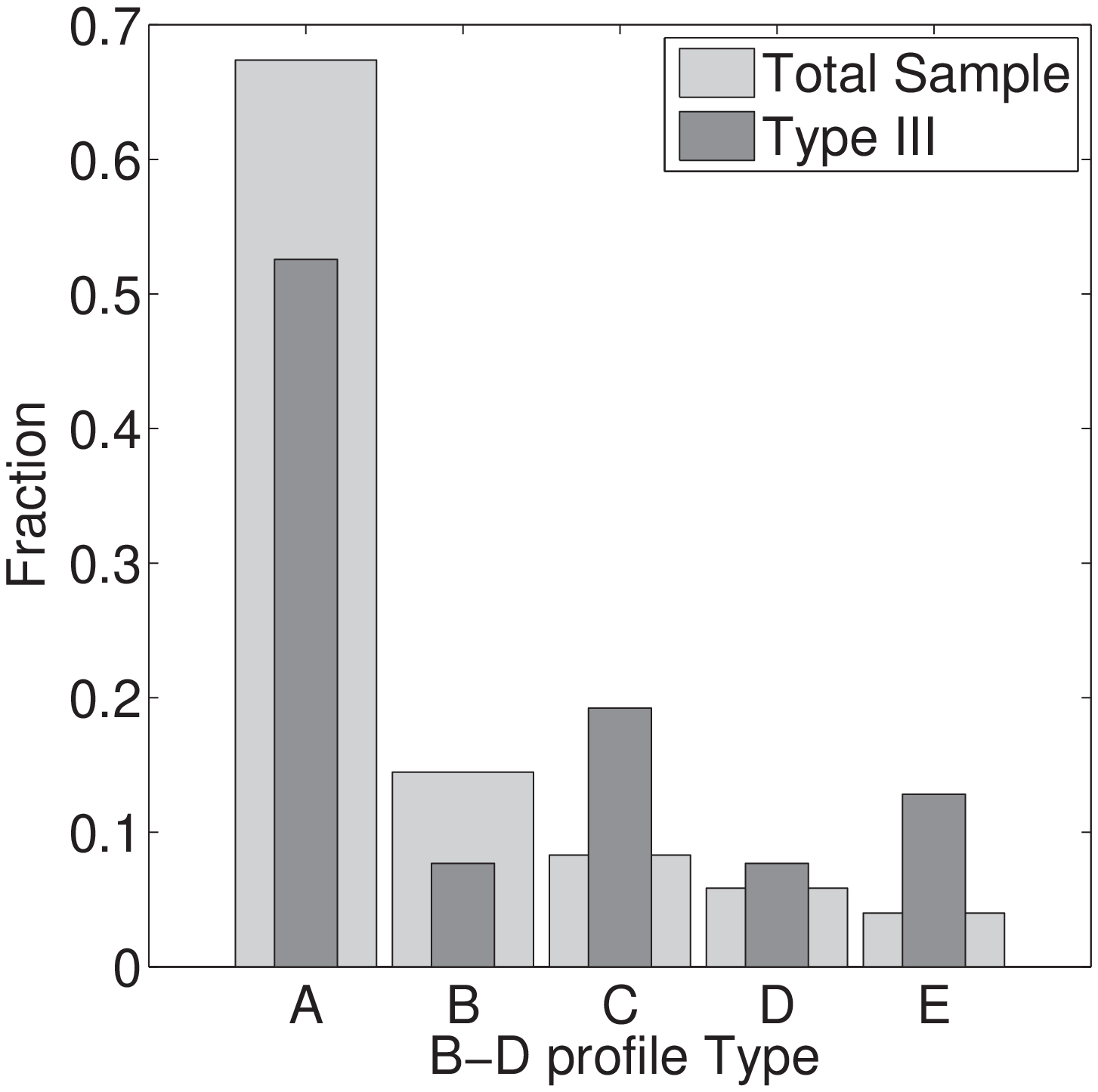}
\caption{\label{B-D profile types} Bulge-disc (B-D) profile types. Top left: Type~A, `classical' profile.
Top right: Type~B, disc-dominated. Middle left: Type~C, bulge-dominated at small/large radii but
disc-dominated at intermediate radii. Middle right: Type~D, bulge-dominated. Bottom left: Type~E,
`constrained' outer bulge. Bottom right: profile type distributions for the total sample and Type~III
sub-sample.}
\end{figure}

B-D decompositions using a de Vaucouleurs ($r^{1/4}$) bulge plus an exponential disc can be classified into
$4$ distinct profile types \citep[e.g.][]{Allen_etal:2006}, see Fig.~\ref{B-D profile types}.

\begin{enumerate}

\renewcommand{\theenumi}{(\arabic{enumi})}
\item {\em Type A:} `Classical' system. The bulge profile dominates at the centre, while the disc profile
dominates at large radii. The bulge/disc profiles cross only once.
\item {\em Type B:} Disc-dominated system. The disc profile dominates at all radii, with a weak contribution
from the bulge in the centre. The bulge/disc profiles never cross.
\item {\em Type C:} The bulge profile dominates at small/large radii, but the disc profile dominates at
intermediate radii.
\item {\em Type D:} Bulge-dominated system. The bulge profile dominates at all radii with a weak underlying
disc component. The bulge/disc profiles never cross\footnote{Note the use of capital letters in profile
types to avoid confusion with other classification schemes.}.
\end{enumerate}

In addition to this, we also observe decompositions where the disc profile dominates in the centre, while
the bulge profile dominates at large radii (hereafter Type E). Here, we believe an outer antitruncated disk
has incorrectly affected the bulge profile fit. In these cases, B-D decomposition is not a true
representation of the galaxy at large radii and these galaxies probably have Type B-like compositions. 
Similar constraints may also occur in some Type~D profiles. 

Fig.~\ref{B-D profile types} shows the distribution of B-D profile types for both the total sample and
the Type III sub-sample. As expected, the fraction of Type~C/E profiles is greater in the Type~III
sub-sample. For the total sample, several other correlations between B-D profile type, Hubble-type
morphology (Sa-Sd) and measured $B/D$ ratio are observed. Type A/B profiles are equally probable in all
Hubble types but Type C/D profiles are more common in earlier Hubble types (Sa-Sb). Also as expected,
mean/median $B/D$ decreases with progressively later Hubble types and increases for the sequence of B-D
profile types B--A--C--D (increasing bulge dominance).

\begin{figure*}
\centering
\includegraphics[width=0.24\textwidth]{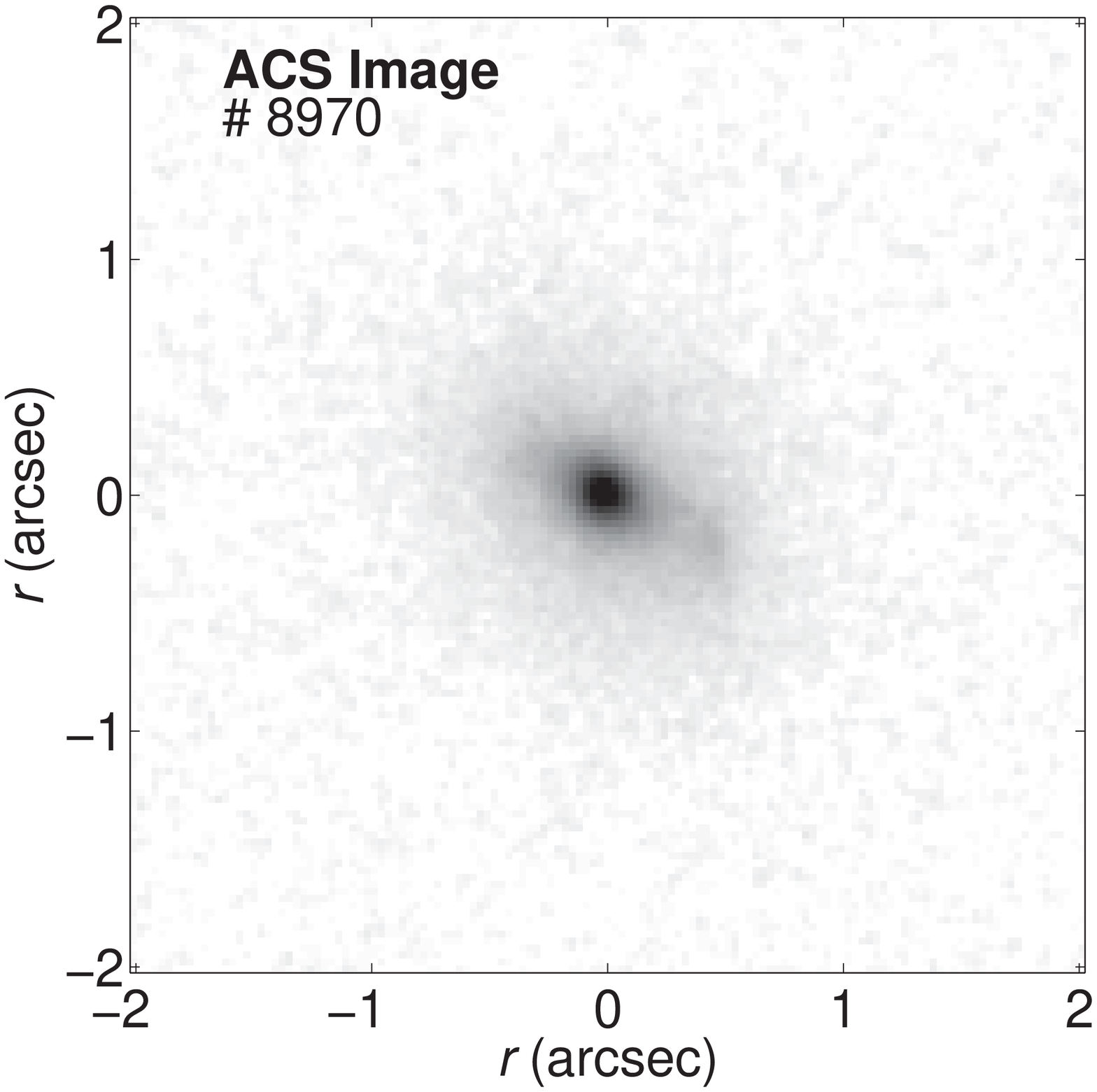}
\includegraphics[width=0.215\textwidth]{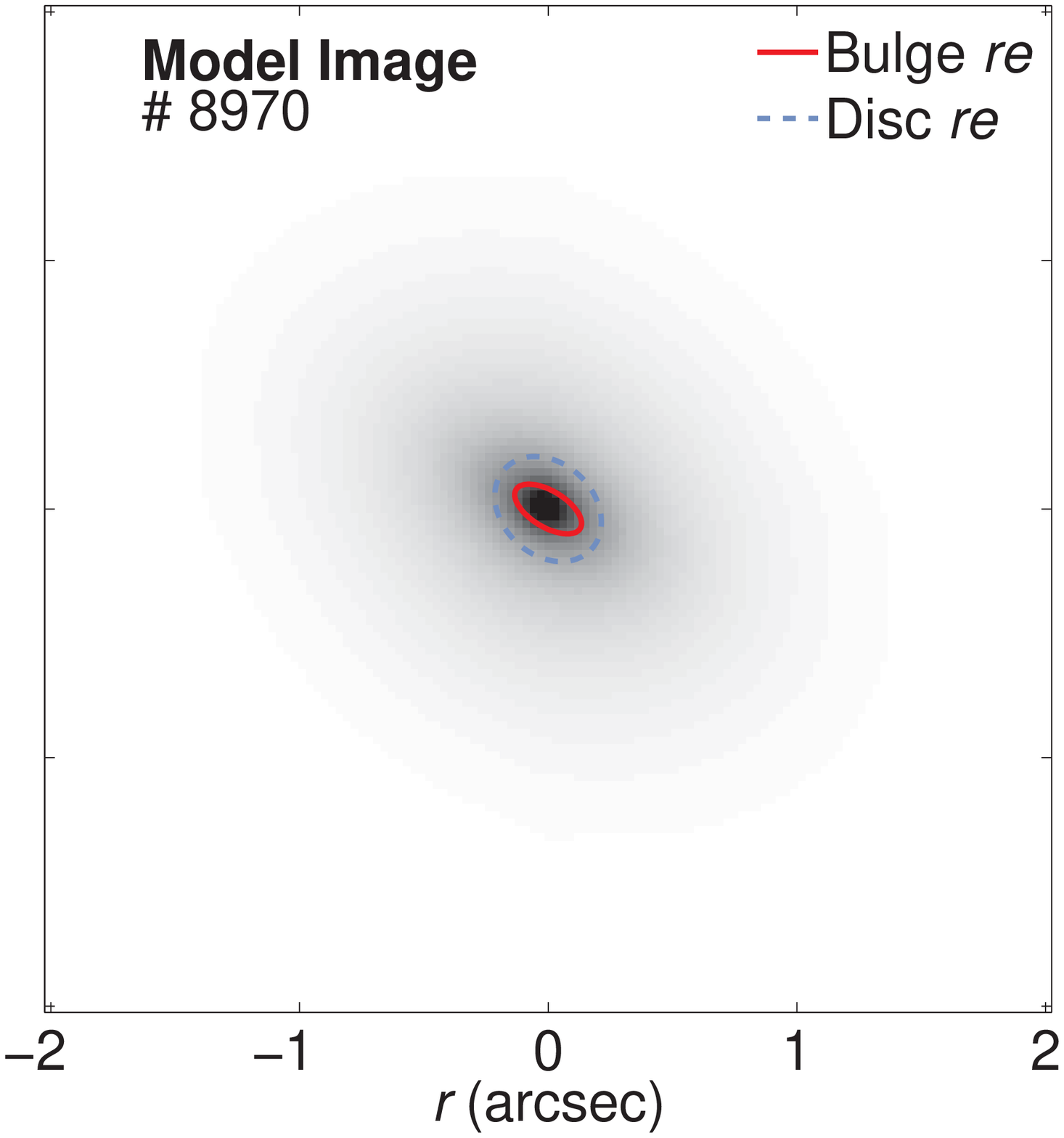}
\includegraphics[width=0.215\textwidth]{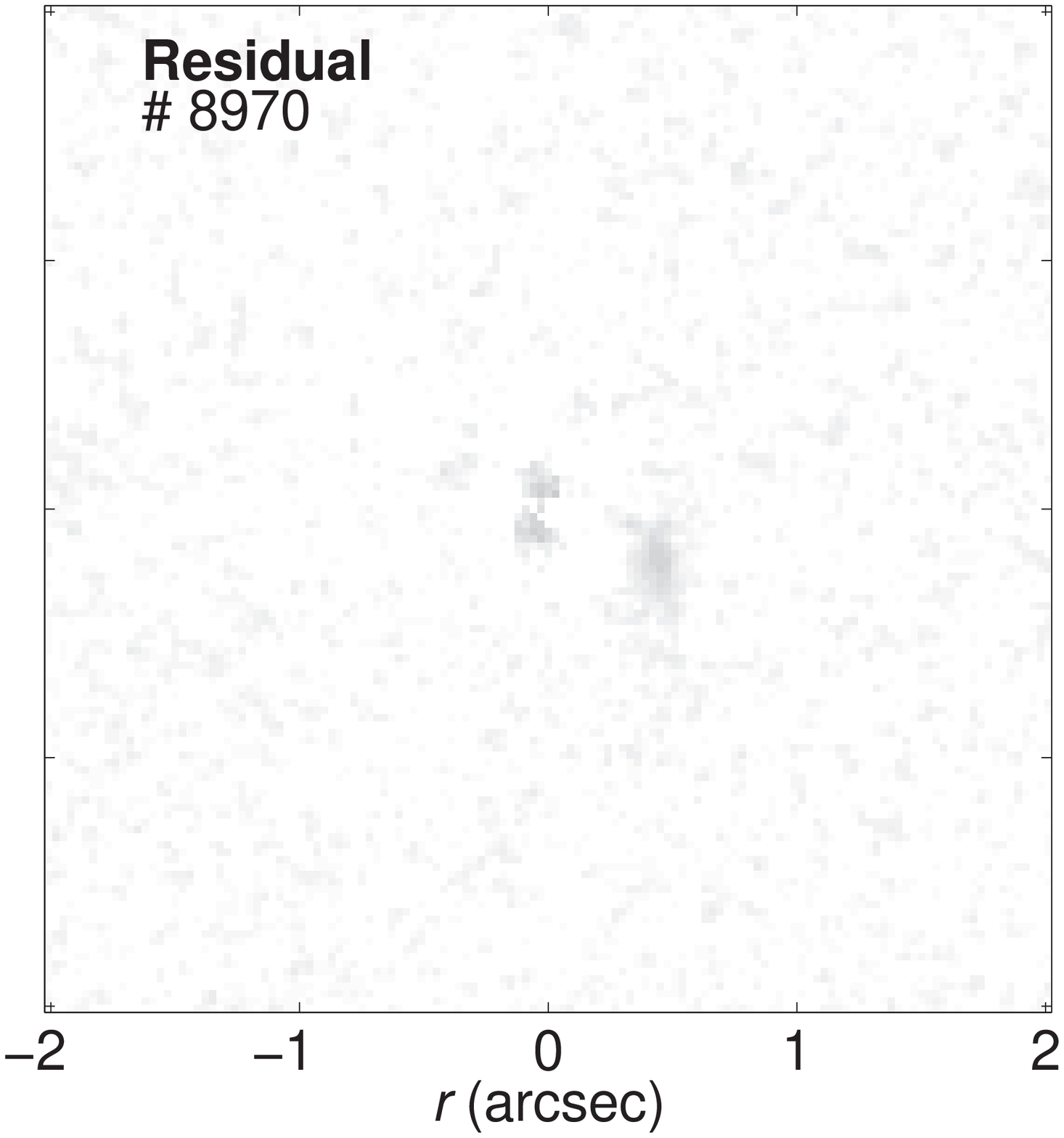}
\includegraphics[width=0.29\textwidth]{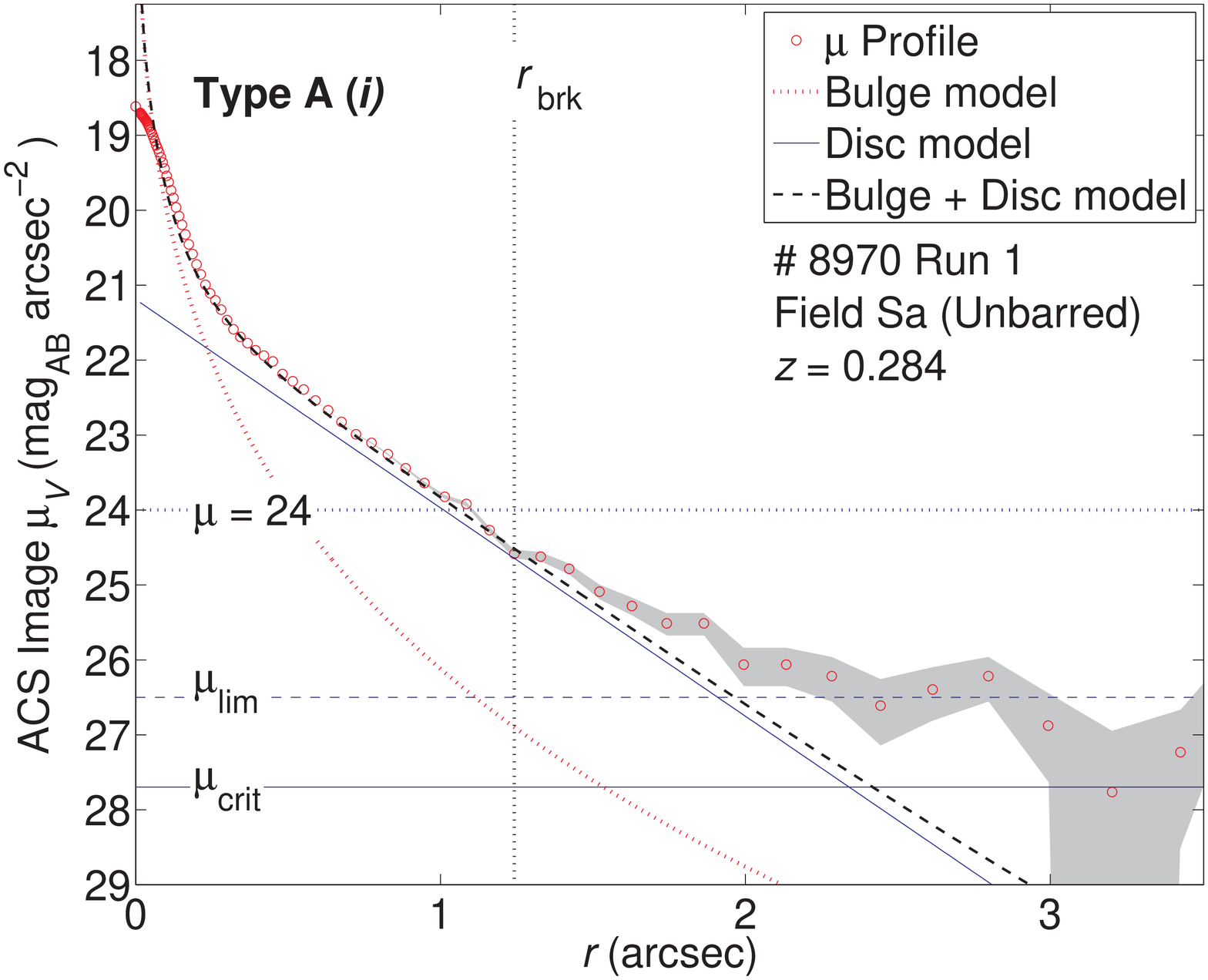}\\
\includegraphics[width=0.24\textwidth]{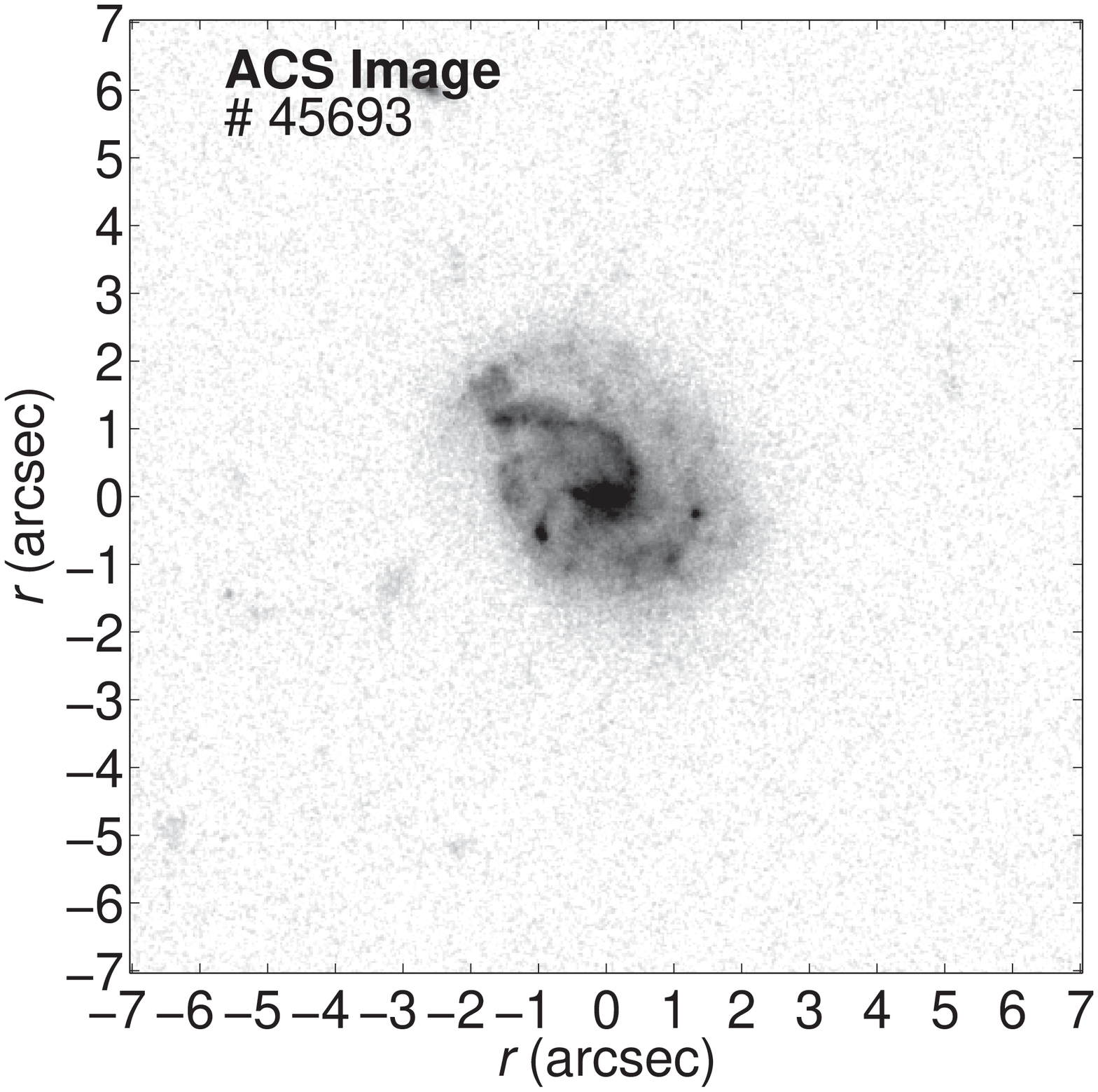}
\includegraphics[width=0.215\textwidth]{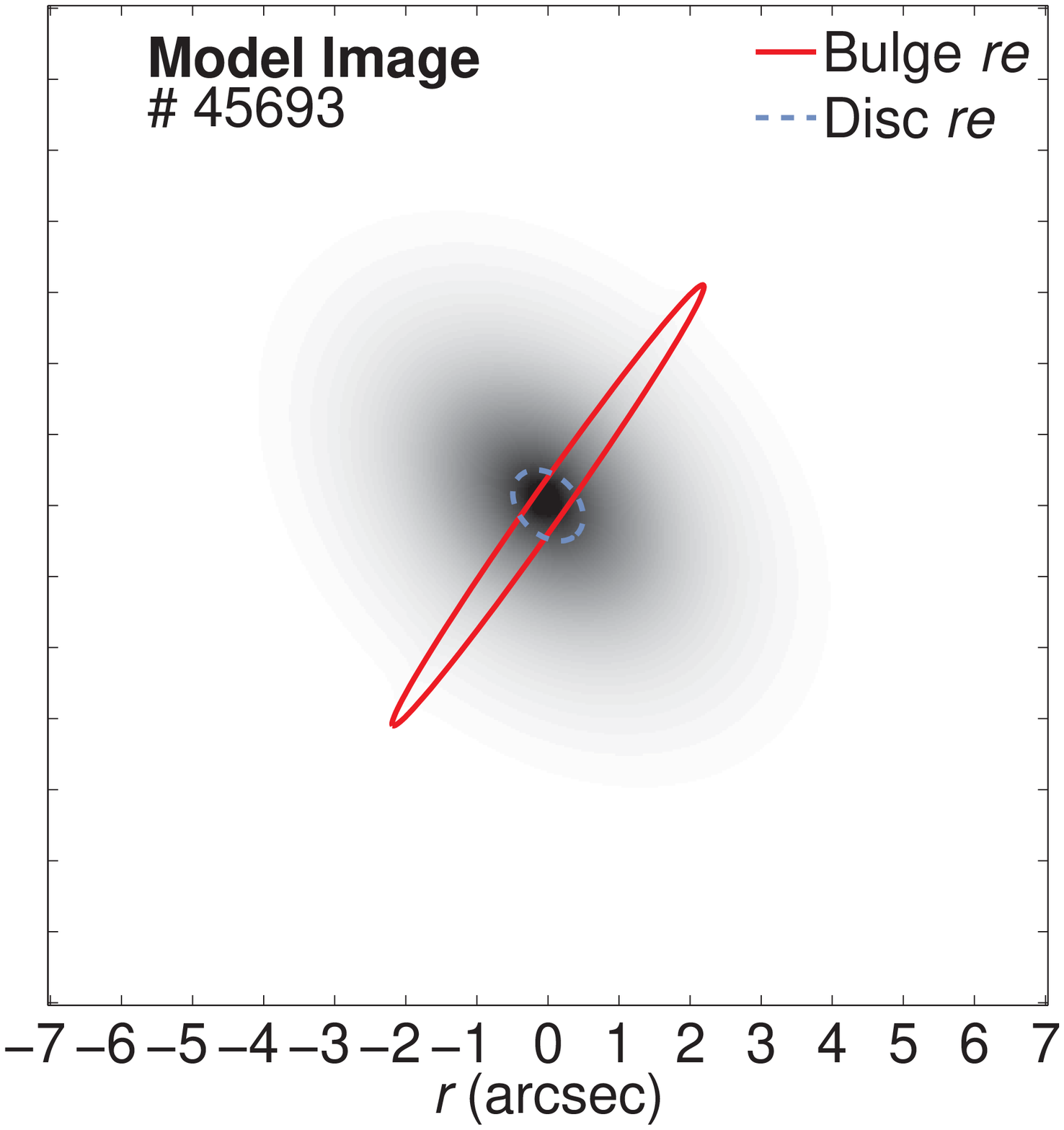}
\includegraphics[width=0.215\textwidth]{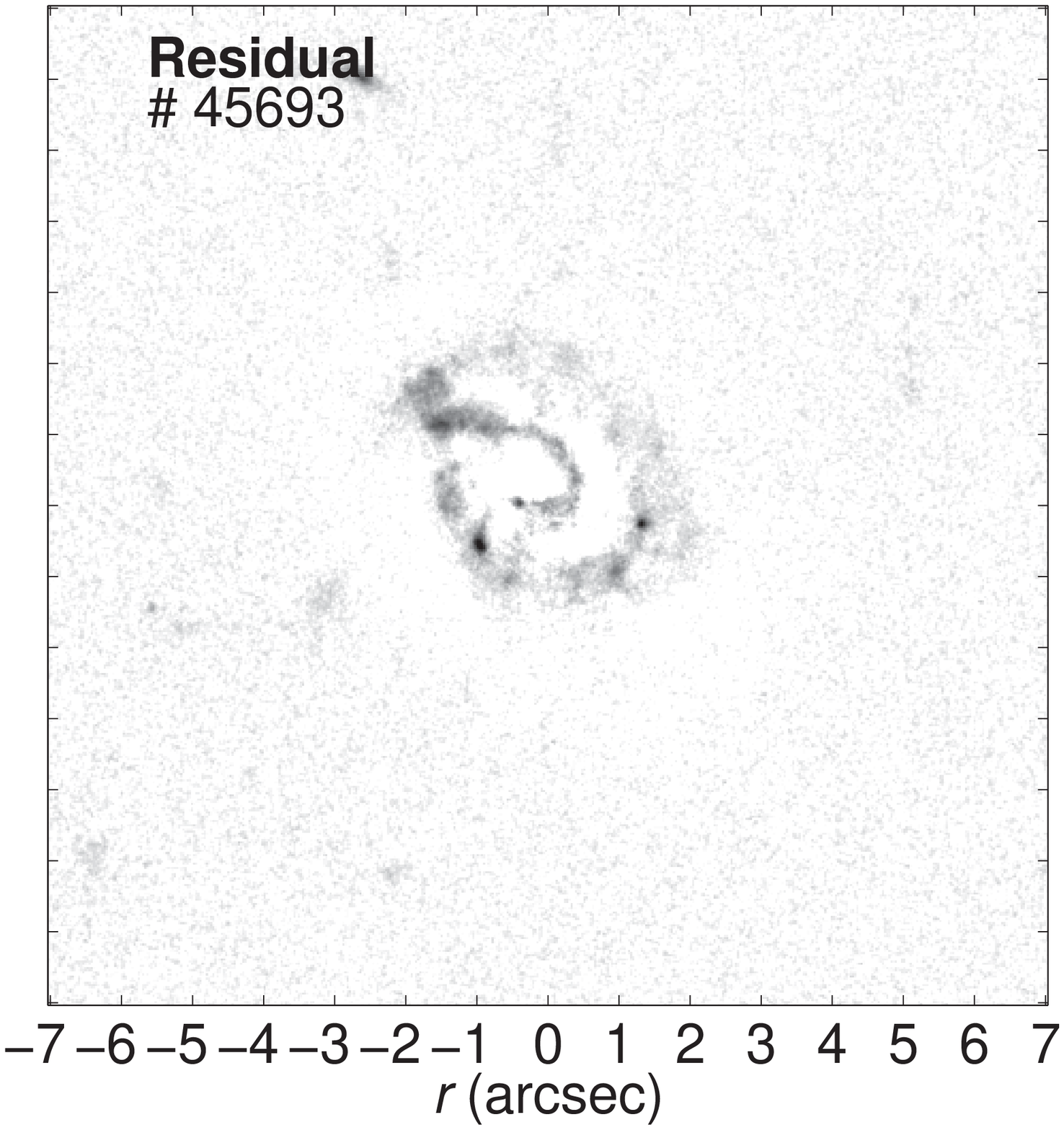}
\includegraphics[width=0.29\textwidth]{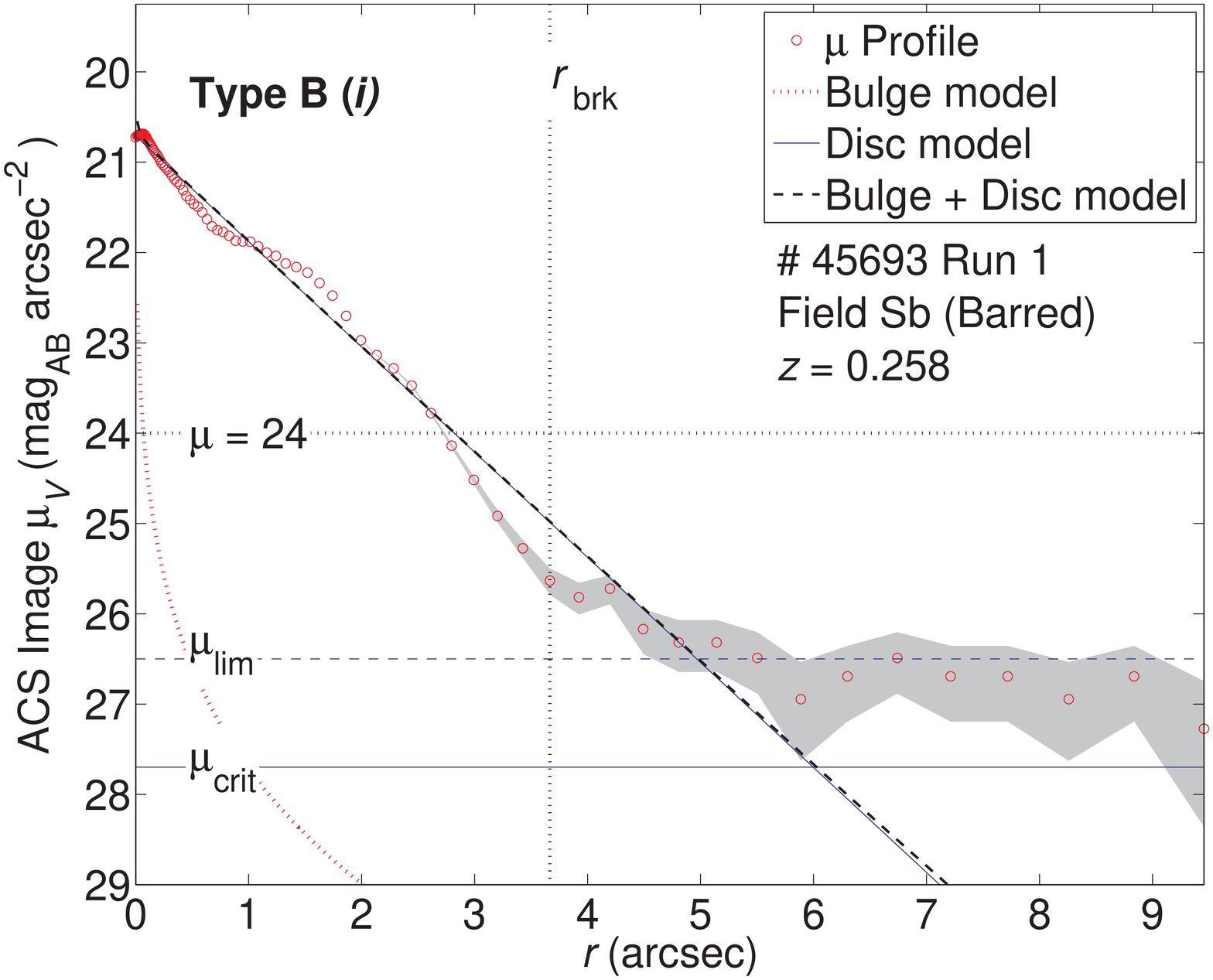}\\
\includegraphics[width=0.24\textwidth]{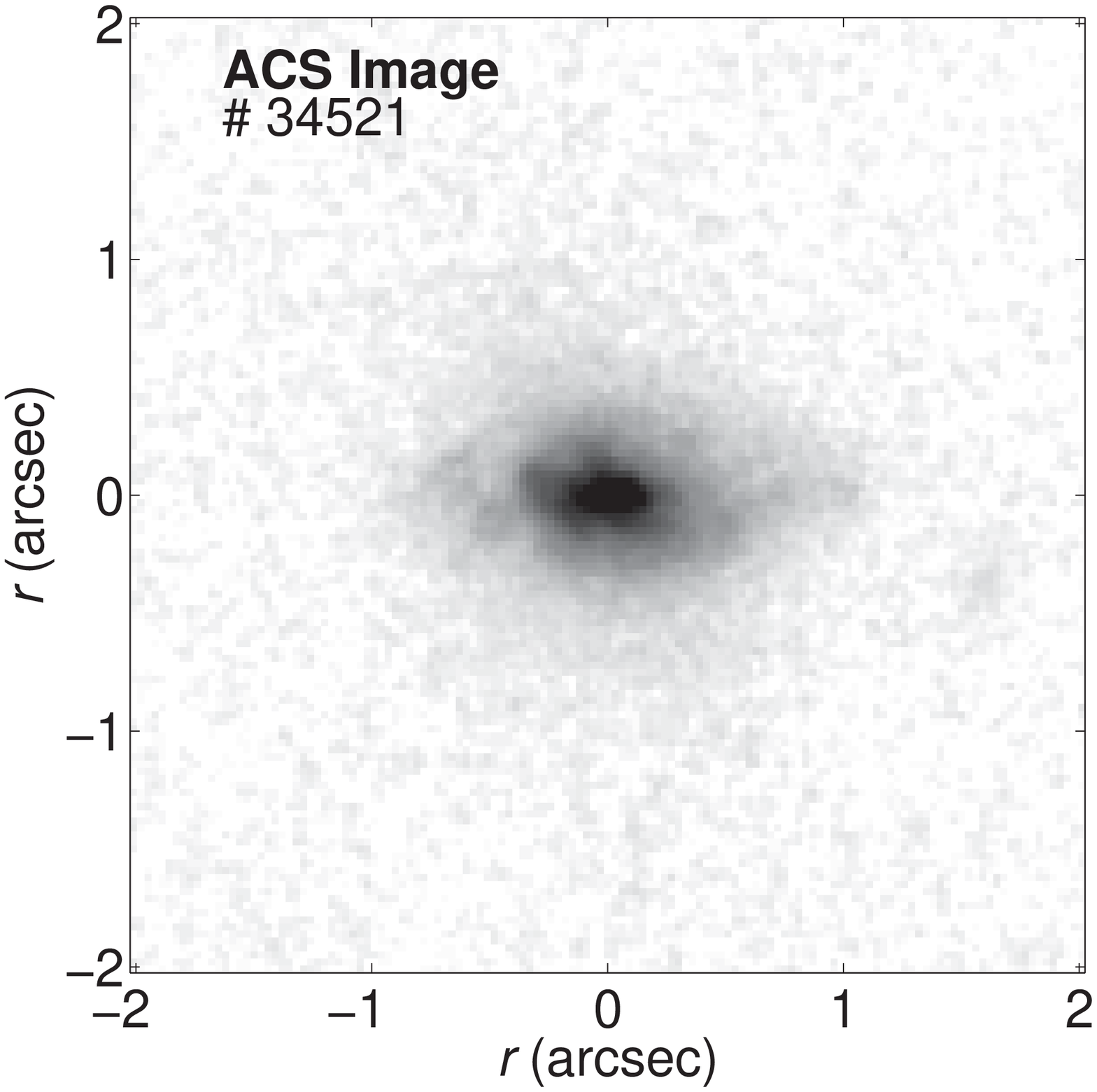}
\includegraphics[width=0.215\textwidth]{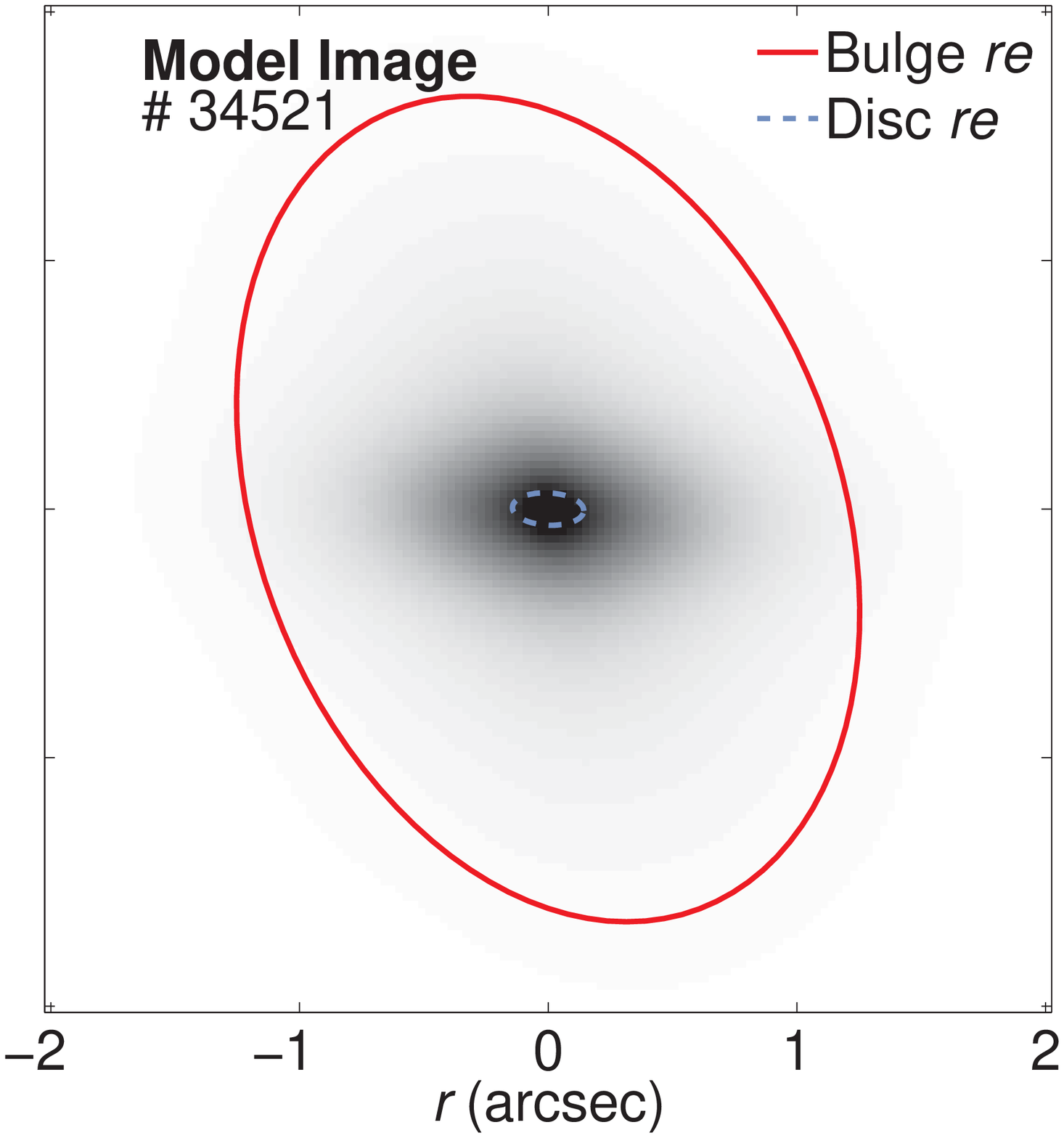}
\includegraphics[width=0.215\textwidth]{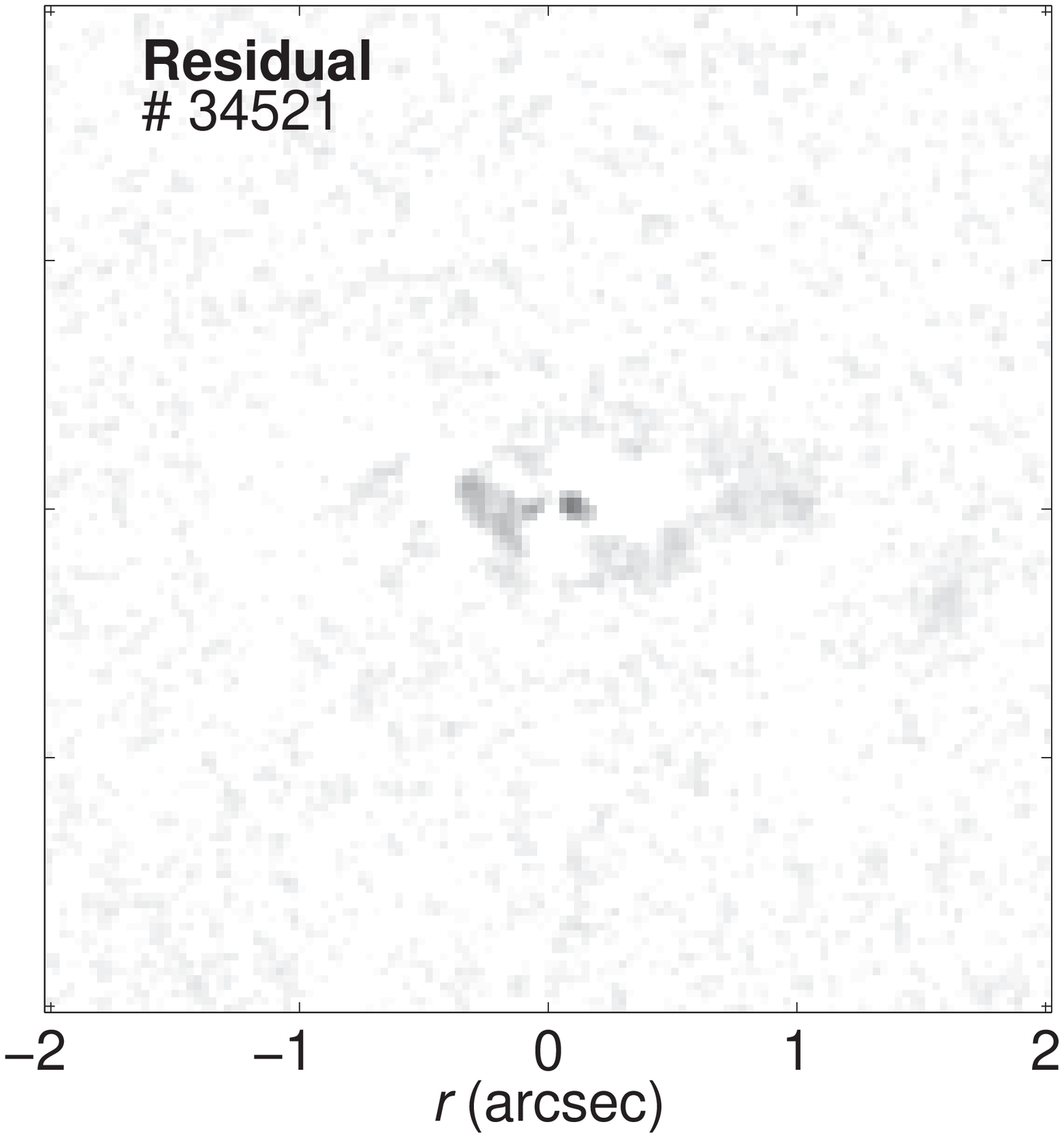}
\includegraphics[width=0.29\textwidth]{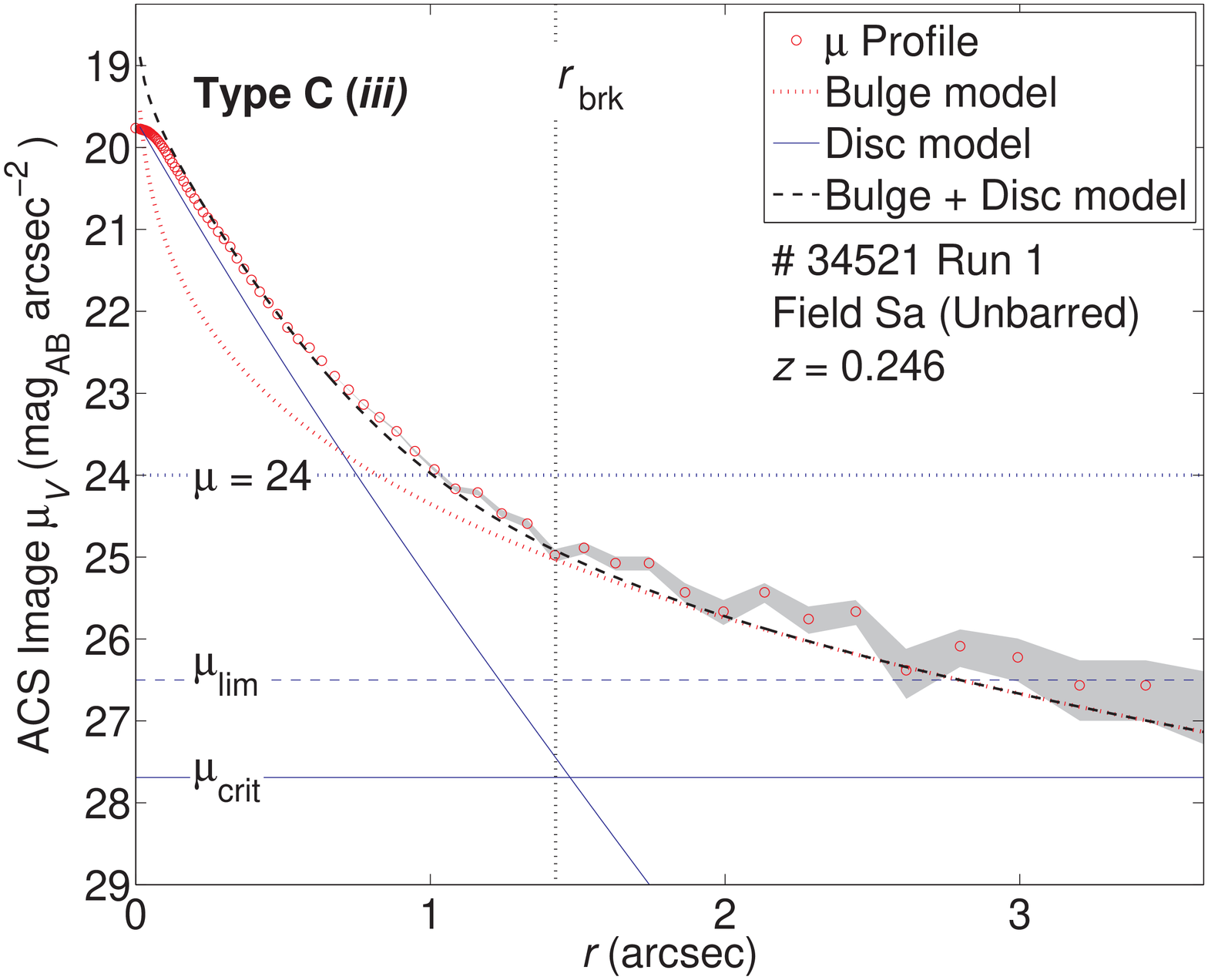}\\
\includegraphics[width=0.24\textwidth]{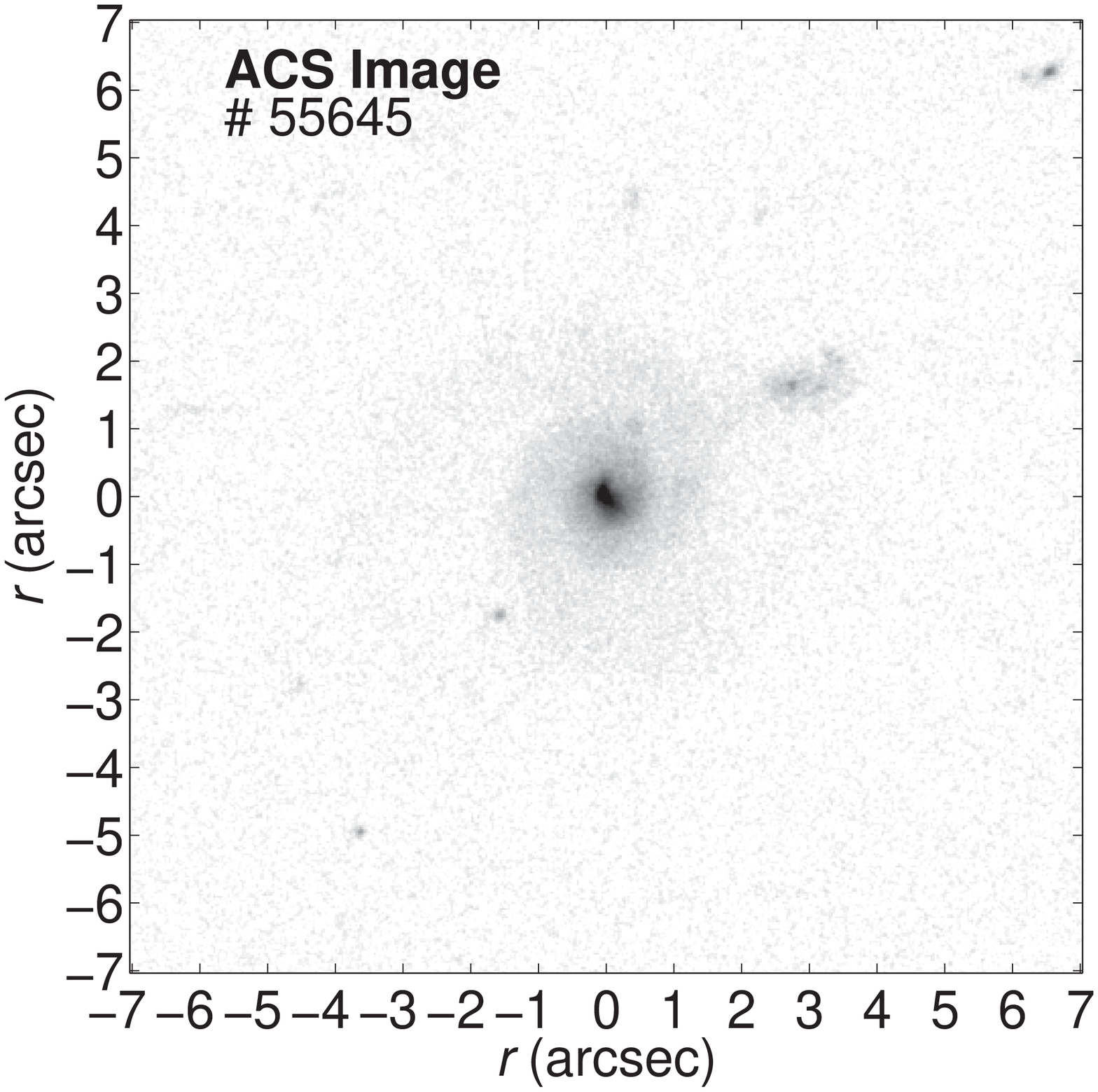}
\includegraphics[width=0.215\textwidth]{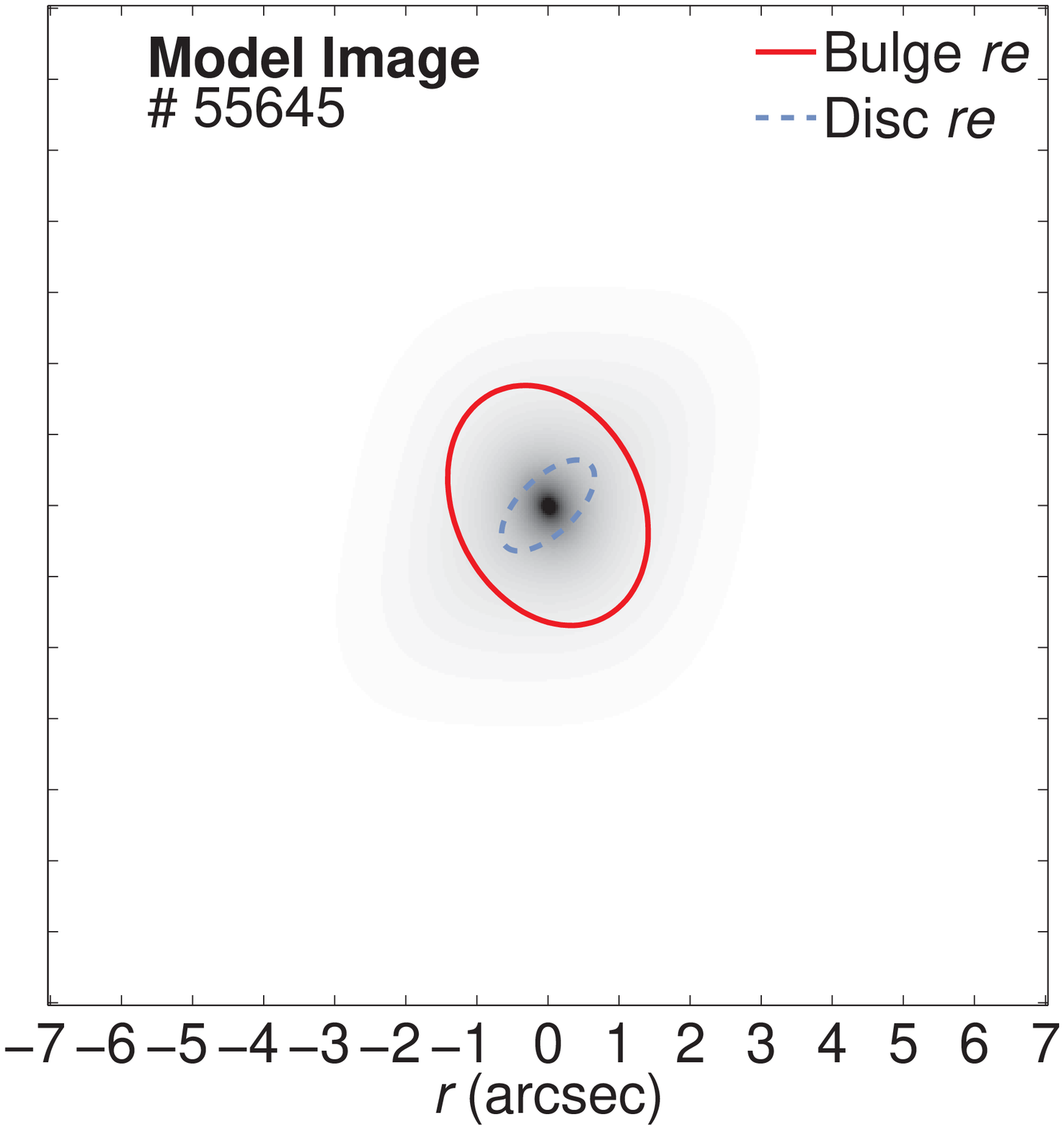}
\includegraphics[width=0.215\textwidth]{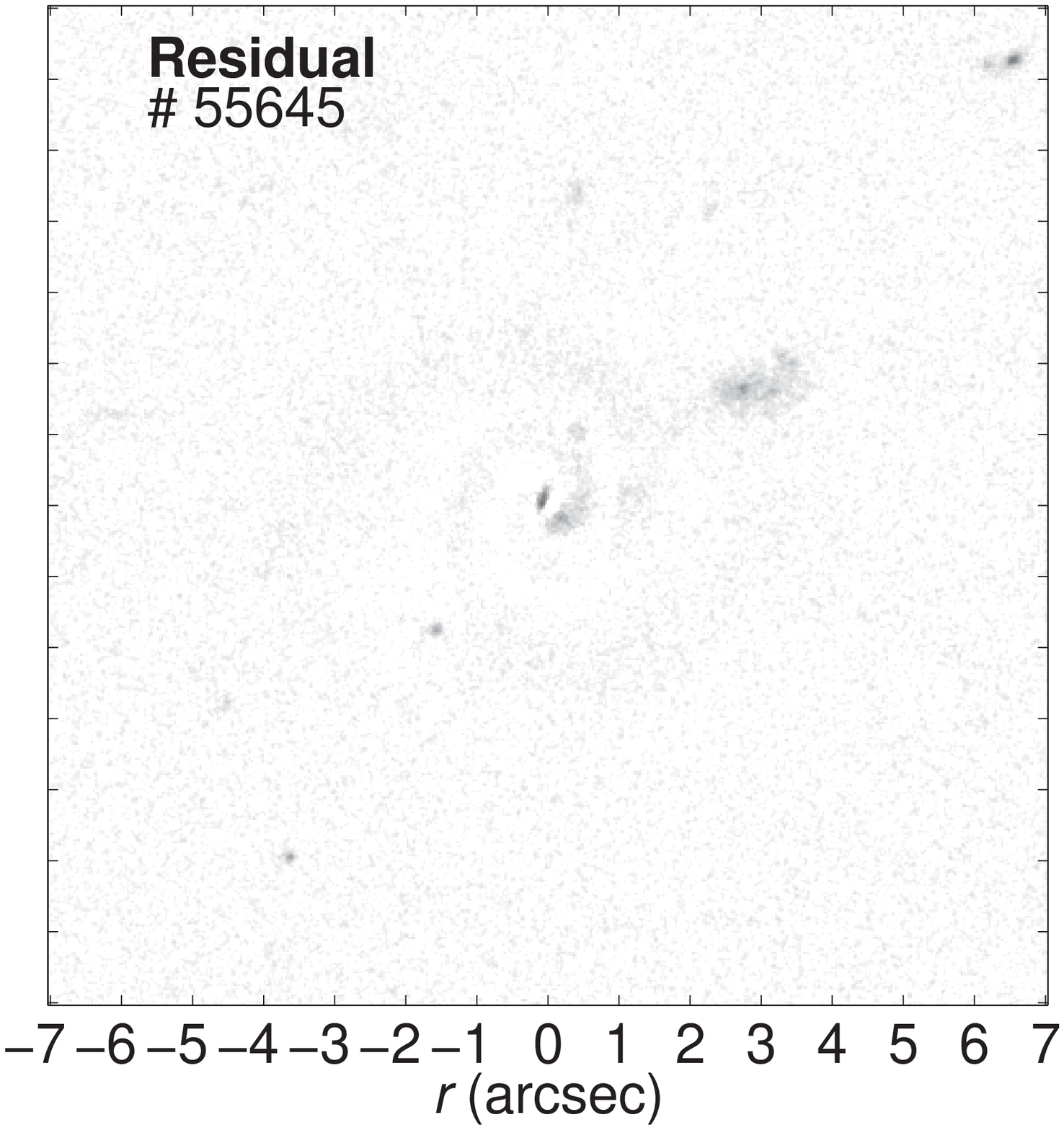}
\includegraphics[width=0.29\textwidth]{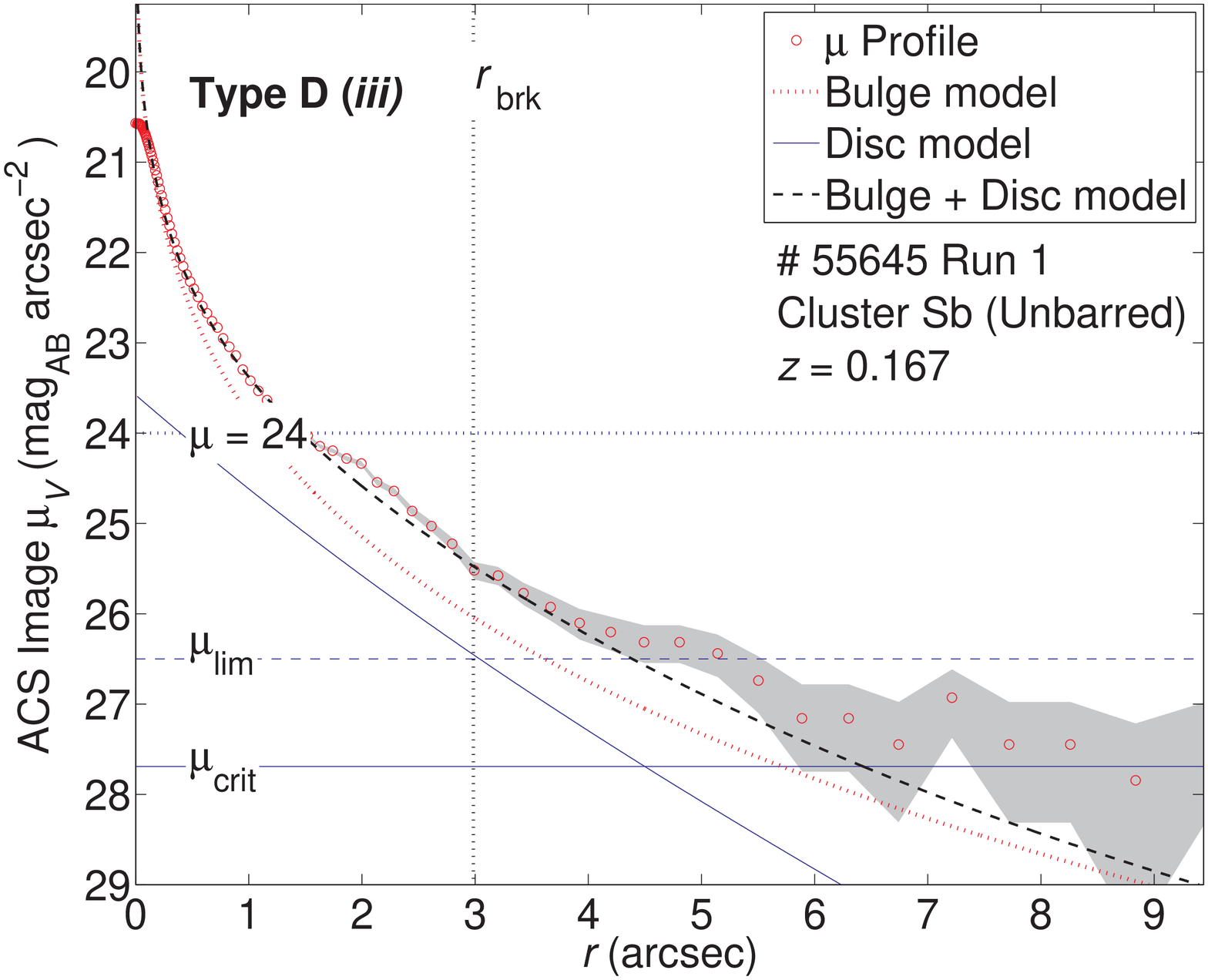}\\
\includegraphics[width=0.24\textwidth]{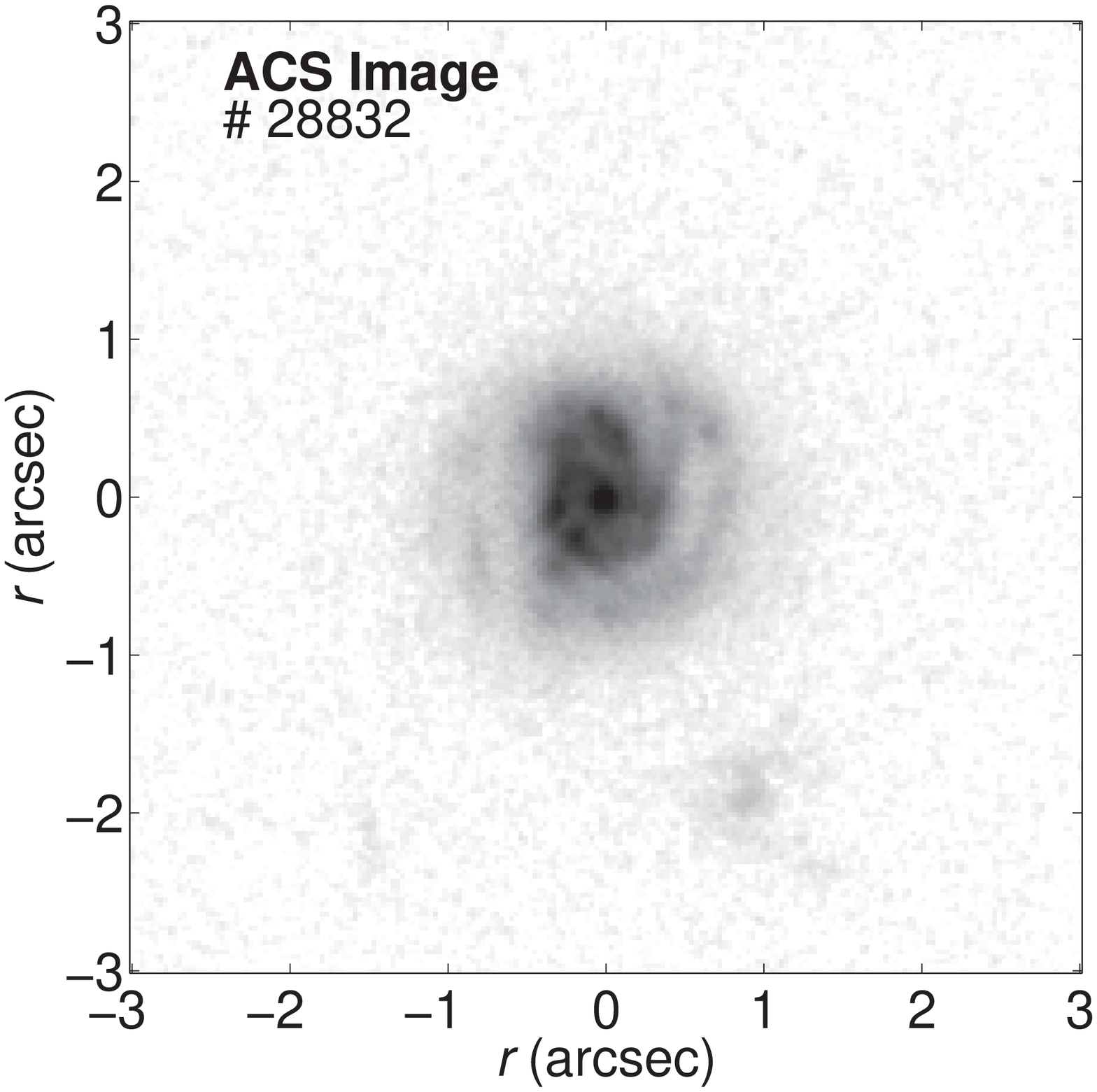}
\includegraphics[width=0.215\textwidth]{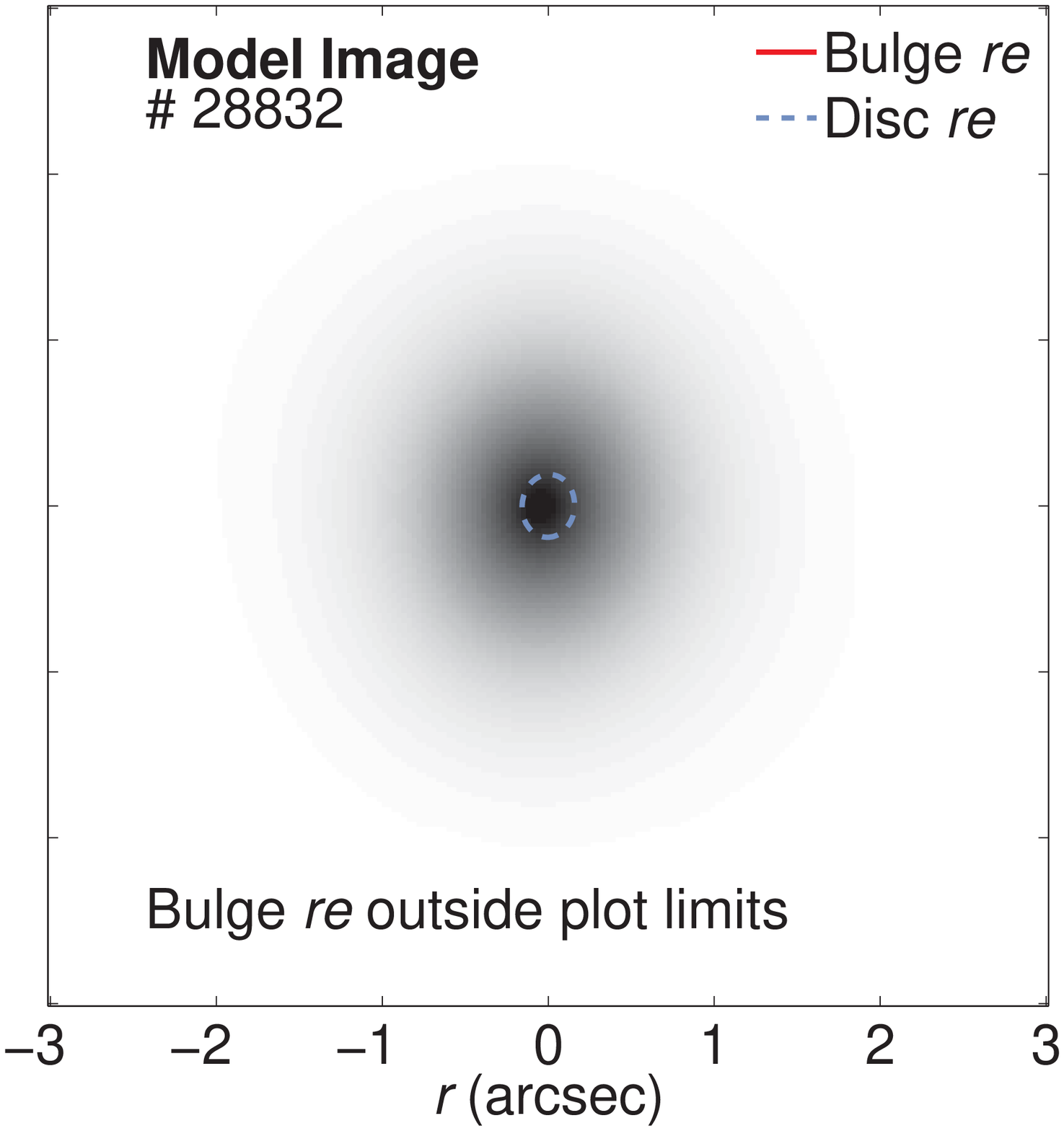}
\includegraphics[width=0.215\textwidth]{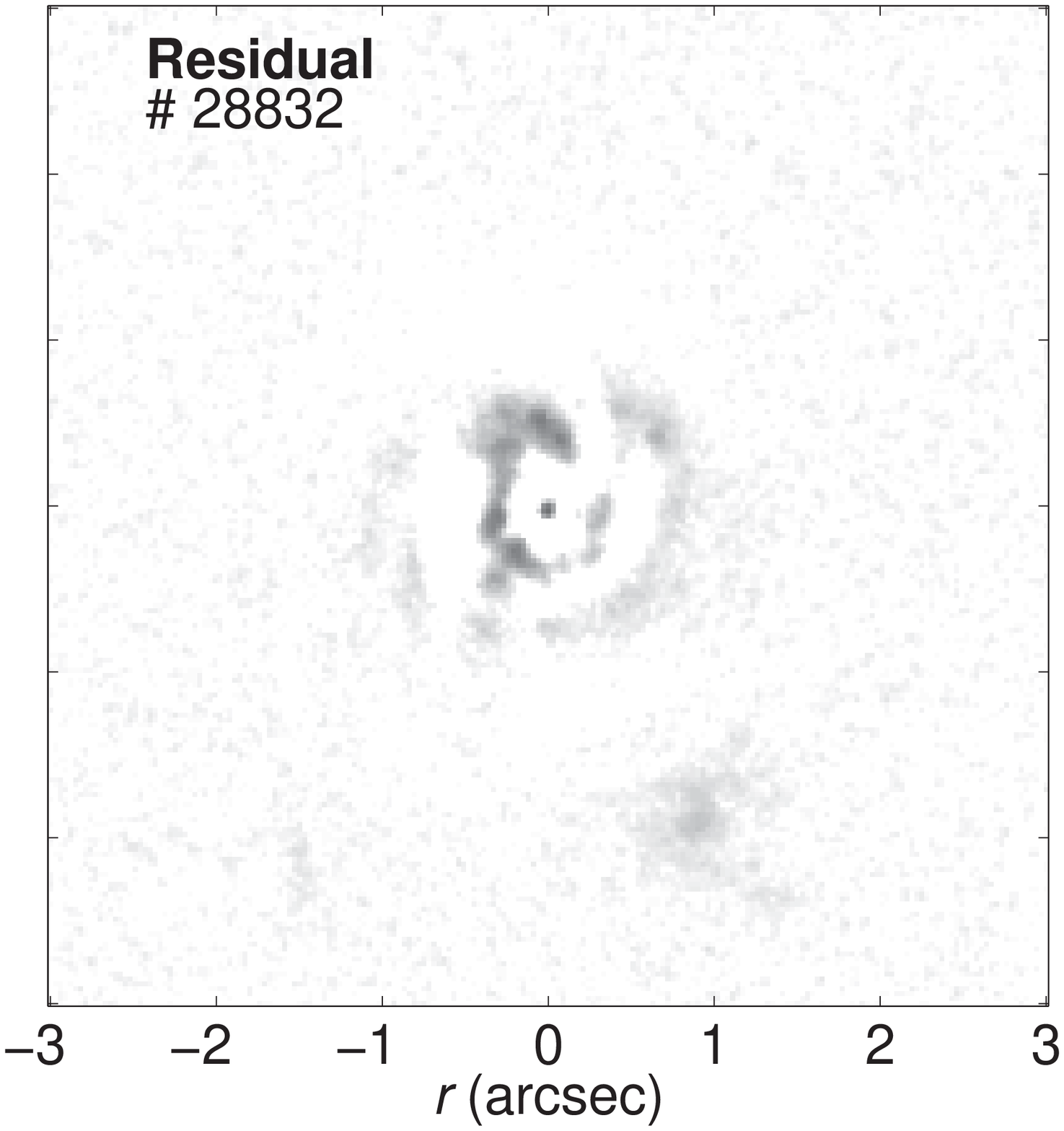}
\includegraphics[width=0.29\textwidth]{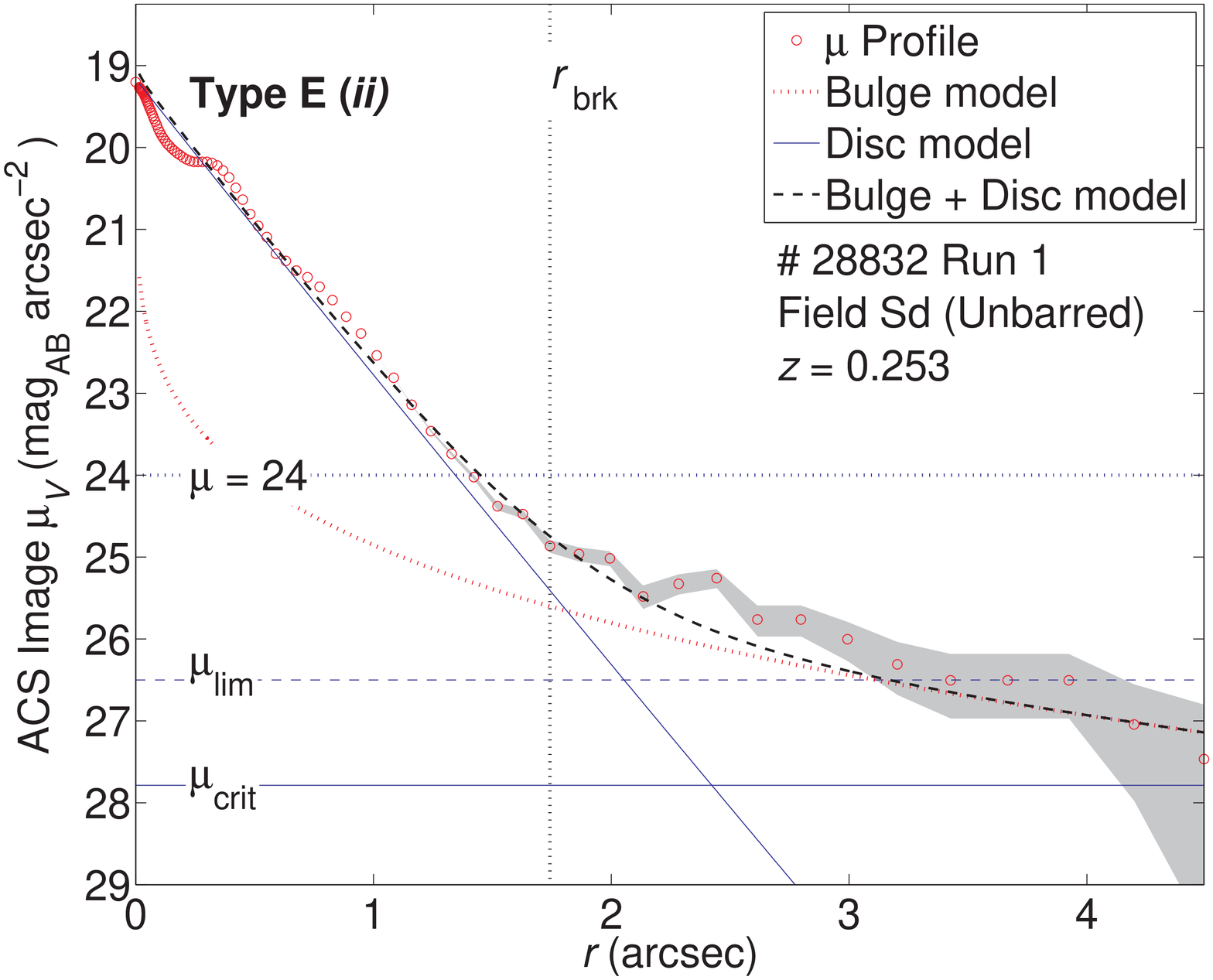}\\

\caption{\label{Examples} Example B-D decompositions and antitruncated $\mu$ profiles. Rows, top to bottom:
decompositions producing Type~A, Type~B, Type~C, Type~D and Type~E profiles. The bulge ({\em red dotted line}),
disc ({\em blue line}), and bulge $+$ disc ({\em black dashed line}) profiles from B-D decomposition are
overplotted on the measured $\mu$ profiles ({\em red circles}). The {\em{i}}/{\em{ii}}/{\em{iii}} notation
after the profile type indicates whether the bulge profile has a negligible, minor or major contribution
outside the break radius $r>r_{\rm brk}$, as in Section~\ref{Results}. Errors in the measured $\mu$ profiles
are for an over- and undersubtraction of the sky by $\pm1\sigma$. The $\mu_{\rm crit}$/$\mu_{\rm lim}$ levels
represent $+1\sigma$/$+3\sigma$ above the sky respectively, see \protect\cite{Maltby_etal:2011} for full
details. The model images show the effective radius isophote for both the bulge ({\em red line}) and disc
({\em blue dashed line}) models.}
\end{figure*}

\begin{figure}
\centering
\includegraphics[width=0.44\textwidth]{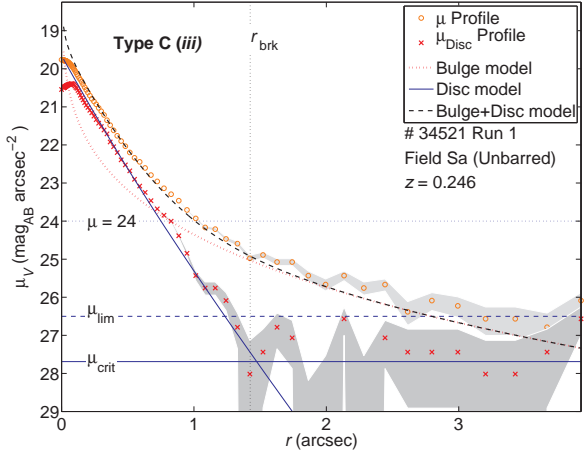}
\caption{\label{Good_example} A rare/unusual example of a bulge profile causing an antitruncation in a $\mu$
profile, (Type~C example from Fig.~\ref{Examples}, break radius $r_{\rm brk}$). The bulge ({\em red dotted
line}), disc ({\em blue line}), and bulge $+$ disc ({\em black dashed line}) profiles from B-D decomposition
are overplotted on the measured $\mu$ profile ({\em red circles}). The disc-residual $\mu_{\rm Disc}$ profile
(measured $\mu$ profile minus bulge-only model, {\em red crosses}) shows no antitruncation (for errors,
see Fig.~\ref{Examples}).}
\end{figure}

For each galaxy in our Type~III sub-sample, we compared the measured, fixed ellipse $\mu(r)$ profile
with the model $\mu$ profiles from the B-D model in order to assess the contribution of bulge light in the
outer regions of the galaxy ($r > r_{\rm brk}$). Fig.~\ref{Examples} shows some examples for each B-D profile
type. These comparisons resulted in three possible scenarios. Bulge light in the outer profile
($r>r_{\rm brk}$) either had:

\begin{enumerate}

\item{\em little or no contribution} ($\sim70$ per cent): For all Type~A/B profiles the bulge contributes
virtually no light at $r > r_{\rm brk}$ and in some Type~C/E profiles the contribution is negligible. This
can be determined by inspection of the measured disc-residual profile $\mu_{\rm Disc}(r)$ and assessing if
the properties of the outer profile/break ($r_{\rm brk}$, $\mu_{\rm brk}$, scalelength) have been affected
with respect to the sky-subtraction error. No Type~D profiles are in this category.

\item{\em minor contribution} ($\sim15$ per cent): Approximately half of these cases are Type~C profiles
where the bulge profile emerges from the end of the disc and contributes some light to the outer regions
of the galaxy. The remaining half are Type~E profiles where the bulge appears to be constrained by an outer
exponential disc. The amount contributed is enough to affect the outer profile causing $\mu_{\rm brk}$ and
the outer scalelength to be different in the disc-residual profile $\mu_{\rm Disc}(r)$. However, the
antitruncation remains present.

\item{\em major contribution} ($\sim15$ per cent): Here, the bulge contributes the majority of the light at
$r > r_{\rm brk}$. Approximately half of these cases are Type~D profiles where the bulge dominates at all
radii. For these cases, the de Vaucouleurs profile is either interpreted as an antitruncation or being
constrained by an outer shallow disc. The remaining half are all Type~C profiles (with one exception which
is Type~E). In these latter cases the antitruncation is not observed in the disc-residual profile
$\mu_{\rm Disc}(r)$, see Fig.~\ref{Good_example} for such an example.

\end{enumerate}

\section[]{Conclusions}

\label{Conclusions}

Our results suggest that for the majority of Type~III profiles ($\sim85$ per cent), the excess light beyond
the break radius $r_{\rm brk}$ is related to an outer shallow disc (Type~III-d). However, it is important to
note that for some of these cases ($\sim15$ per cent), bulge light can affect the properties of the disc
profile (e.g. $\mu_{\rm brk}$, outer scalelength). For the remaining Type~III profiles ($\sim15$ per cent),
the excess light at $r>r_{\rm brk}$ can be attributed to the bulge profile (Type~III-s). However, few of
these latter cases (only 3 galaxies with stable decompositions, $\sim5$ per cent) exhibit profiles where the
bulge profile extends beyond a dominant disc (Type~C) and causes an antitruncation in the $\mu$ profile.

Previous works, \citep{Erwin_etal:2008, Gutierrez_etal:2011} have used the ellipse method to classify their
Type~III-d/III-s profiles. The results of these works are in good agreement with their Type~III profiles
being $\sim60$ per cent Type~III-d and $\sim40$ per cent Type~III-s. These results appear {\em not} to
be consistent with our results. However, consider the case where the contribution of bulge light at
$r>r_{\rm brk}$ is too little to explain the antitruncation but great enough to affect the properties of the
profile. The increased contribution of bulge light to the $\mu(r)$ profile at $r_{\rm brk}$ could lead to a
smoothing of the inflection and a roundening of the outer isophotes. Therefore, using the ellipse method
would likely have resulted in these Type~III-d profiles being classified as Type~III-s. Applying this
reasoning to our results, we obtain fractions of $\sim70$/$30$ per cent for Type~III-d/III-s profiles
respectively. Therefore, the ellipse method potentially could lead to genuine disc breaks being classified
as \mbox{Type~III-s}. Our method offers an improved way to determine whether an antitruncation is disc or
spheroid related.

However, our method does have some obvious drawbacks. In a two-component B-D decomposition, an outer
antitruncated disc could cause the bulge profile to be constrained and lead to an over-estimation of bulge
light in the outer regions of the galaxy. This naturally enhances our fraction of Type~III-s profiles which
therefore represents an upper limit to the fraction of genuine Type~III-s profiles in our Type~III sample.
Therefore, we can conclude that in the vast majority of cases Type~III profiles are indeed a true disc
phenomenon.

Several studies have proposed formation scenarios for Type~III profiles, mainly via satellite accretion or
minor mergers \citep{Penarrubia_etal:2006, Younger_etal:2007}. A discussion of these scenarios is beyond the
scope of this work. However, we do make one important comment. Previous works
\citep[e.g.][]{Pohlen_Trujillo:2006, Maltby_etal:2011} have reported that Type~III profiles are more
frequent in earlier Hubble types. This result is consistent with a minor merger scenario for their origin.
However, the excess Type~III profiles in early-types could easily have been related to a natural increase in
the number of Type III-s profiles in earlier Hubble types. Fortunately, we observe exactly the same relations
using just our genuine Type~III-d profiles and thus our results remain consistent with the minor merger
scenario for the formation of Type~III profiles.

\section[]{Acknowledgements}

We wish to thank the anonymous referee for their very careful reading and detailed comments on the original
version of this paper which have helped us to vastly improve it. We also thank Stephen Bamford for useful
discussions. The support for STAGES was provided by NASA through GO-10395 from STScI operated by AURA under
NAS5-26555. DTM and MEG were supported by STFC. CH acknowledges support from a Spanish MEC postdoctoral
fellowship.


\bibliographystyle{mn2e} \bibliography{DTM_bibtex} \bsp

\label{lastpage}

\end{document}